\documentstyle[12pt,aps,epsf,epsfig]{revtex}
\newcommand{\comp}{{\rm C}\hspace{-1ex}\rule{0.1mm}{1.5ex}\hspace{1ex}}
\newcommand{\reales}{{\rm R}\hspace{-1ex}\rule{0.1mm}{1.5ex}\hspace{1ex}}

\def\er#1#2{\relax\ifmmode{}^{+#1}_{-#2}\else$^{+#1}_{-#2}$\fi}
\newcommand{\be}{\begin{equation}}
\newcommand{\bea}{\begin{eqnarray}}
\newcommand{\ee}{\end{equation}}
\newcommand{\eea}{\end{eqnarray}}

\def\({\Big(}
\def\){\Big)}
\def\slashchar#1{\setbox0=\hbox{$#1$}
   \dimen0=\wd0 \setbox1=\hbox{/} \dimen1=\wd1
   \ifdim\dimen0>\dimen1 \rlap{\hbox to \dimen0{\hfil/\hfil}} #1
   \else  \rlap{\hbox to \dimen1{\hfil$#1$\hfil}} / \fi}
\def\P{\slashchar{P}}
\begin{document}
\draft \tighten \def\footnoterule{\kern-3pt \hrule width\hsize
\kern3pt} \title{The $S_{11}-$ $N$(1535) and $-N$(1650) Resonances in
Meson-Baryon Unitarized Coupled Channel Chiral Perturbation
Theory} \author{J. Nieves\footnote{email:jmnieves@ugr.es} and
E. Ruiz Arriola\footnote{email:earriola@ugr.es} } \address{ {~} \\
Departamento de Fisica Moderna \\ Universidad de Granada \\
E-18071 Granada, Spain }

\date{\today}
\maketitle 

\thispagestyle{empty}

\begin{abstract}

The $s-$wave meson-baryon scattering is analyzed for the strangeness
$S=0$ sector in a Bethe-Salpeter coupled channel formalism
incorporating Chiral Symmetry. Four
channels have been considered: $\pi N$, $\eta N$, $K \Lambda$, $K
\Sigma$. The needed two particle irreducible matrix amplitude  is taken
from lowest order Chiral Perturbation Theory in a relativistic
formalism and low energy constants are fitted to the elastic $\pi N $
phase-shifts and the $\pi^- p \to \eta n$ and $\pi^- p \to K^0 \Lambda$
cross section data. The position of the complex poles in the second
Riemann sheet of the scattering amplitude determine masses and widths of
the $S_{11}-$ $N$(1535) and $-N$(1650) resonances, in reasonable
agreement with experiment. A good overall description of data, from
$\pi N$ threshold up to 2 GeV, is achieved keeping in mind that the
two pion production channel has not been included.

\end{abstract}


\vspace*{1cm}
\centerline{\it PACS: 11.10.St;11.30.Rd; 11.80.Et; 13.75.Lb;
14.40.Cs; 14.40.Aq\\}
\vspace*{1cm} \centerline{\it Keywords: Chiral Perturbation Theory,
Unitarity, $\pi N$-Scattering, } \centerline{\it $S_{11}-N$ Resonances,
Coupled channels, Bethe-Salpeter Equation.}

\newpage
\setcounter{page}{1}

\section{Introduction}

The $N(1535)$ and $N(1650)$ resonances appear as outstanding features
not only in elastic $\pi N$ scattering in the strangeness zero $S_{11}$
($L_{2T,2J}$) partial wave but also in other meson-baryon reactions at
intermediate energies. In quark model approaches these excited nucleon
resonances are mainly composites of three valence quarks, and their
widths are computed as matrix elements of hadronic transition
operators. However, the description of hadron scattering reactions
becomes cumbersome in this framework, and it requires quite elaborated
techniques as the resonating group approach, where it becomes
extremely difficult to impose Chiral Symmetry
(CS)~\cite{GR96}. 

Renouncing to find out a picture of the hadron as a
valence quark bound state, a different point of view consists of
describing scattering reactions taking the hadrons as the relevant
degrees of freedom at low energies. Then, resonances manifest
themselves as poles of the scattering amplitude in a certain Riemann
sheet in the complex energy plane. To perform such a program requires
to implement unitarity in the model. A multichannel $K-$matrix method is
used in the work of Ref.~\cite{VDL00}. Though CS is not incorporated,
this phenomenological approach is able to reproduce a large amount of
data related to the $\pi N \to \pi N$ reaction. The, three body final
state, two pion production channel $\pi\pi N$ is incorporated through
an effective use of two body channels with higher mesonic and baryonic
resonances. In this paper, we will work in this latter type of
approaches, but explicitly imposing CS constraints as an indirect way
of incorporating the bulk of the underlying Quantum Chromo Dynamics
(QCD). Thus, we will establish a unitarity scheme based on the Chiral
Perturbation Theory (ChPT) amplitudes.

CS provides important constraints to the description of low energy
hadronic processes and, in particular, to baryon-meson dynamics. There
have been previous studies of the $\pi N-$ $S_{11}$ partial wave using
a coupled channel formalism and imposing CS constraints. In
Ref.~\cite{KSW95} a Schr\"odinger coupled channel treatment was
employed and the additional inclusion of phenomenological hadronic
form factors was invoked. Within this framework, the $p-$wave
contribution has also been recently examined~\cite{Caro00}.  The two
pion production channel is not considered in these works. Perturbative
estimates~\cite{BKM97a}--\cite{JM97} for the reaction $\pi N \to \pi
\pi N$ indicate that this three-body channel keeps moderately small
not only at threshold but also in a certain region above it. An
attempt to include the two pion production reaction can be found in
Ref.~\cite{BSSN95}, but the treatment of the $\pi \pi N$ channel is
effective and it is represented by a unphysical two body channel which
represents all remaining processes.

In Ref.~\cite{Na00} the Bethe-Salpeter Equation (BSE) has been
employed in the spirit of an Effective Field Theory (EFT). There, the $\pi
\pi N$ channel is  not considered either and the
authors require the introduction of less renormalization constants 
than allowed by CS. Despite of these restrictions, the model
describes not only the elastic $\pi N$
channel, but also the two-body inelastic ones in an  energy
window around the $N(1535)$ resonance. Nevertheless, it fails at
threshold~\cite{private}. Given this partial success and the great
technical difficulties to solve the BSE incorporating the three body
$\pi\pi N$ channel, one might wonder what features of the data can be 
explained incorporating CS constraints and restoring two body unitarity.

 In the present work, we restrict our study to the non-strange
meson-baryon $S_{11}$ partial wave and adopt a similar framework as in
those references, but with some important differences. First, we will
implement exact unitarity by solving the BSE taking the needed input
from lowest order relativistic ChPT. A similar program has been
successfully undertaken both in the pion-pion sector~\cite{EJ99} and
in the $\pi N-$$P_{33}$ partial wave~\cite{EJ01}. Thus, we avoid the
use of phenomenological form-factors and all required information (low
energy constants) can be, in principle, obtained from higher orders in
the chiral expansion. Besides, we aim at describing not only a narrow
energy window, placed at threshold or in the neighborhood of  some
resonance, but also a wider energy region ranging from $\pi N$
threshold up to almost a Center of Mass (CM) meson-baryon energy of
$\sqrt{s}=2$ GeV.

As we discussed at length in Ref.~\cite{EJ99} the BSE, in the context
of EFT's, can be solved in two different schemes: off-shell and
on-shell. Here, we use the off-shell scheme because of the lack of
information on the next-to-leading order in the chiral expansion. In
this scheme, the on-shell scattering amplitude requires some knowledge
of the off-shell behavior of the two particle irreducible amplitude
({\it potential}). After renormalization of the amplitude this
off-shell input leads to a finite number of phenomenological constants
which encode the detailed underlying short-distance dynamics. In
practice, these constants can be either fitted to experiment or
determined by matching the resulting Bethe-Salpeter (BS) amplitude to
standard ChPT\footnote{Ideally, these phenomenological parameters
should be computed from first principles, a yet impossible
task.}. Obviously, the method of determining the constants by matching
to ChPT seems a better one than a direct fit to experimental
data\footnote{In addition, if the matching is possible the off-shell
scheme becomes unnecessarily complicated, as compared to other methods
directly unitarizing the final on-shell amplitude given in terms of
the standard low energy constants. For details, on the on-shell BSE
approach or on the Inverse Amplitude Method (IAM) for meson-meson
scattering, see for instance, the thorough discussion in the second
entry of Ref.~\cite{EJ99}. For $\pi N$ elastic scattering, recent IAM
studies have been pursued in Refs.~\cite{pn00} and~\cite{ej00c}.}. For
the case of meson-baryon scattering, the only known information coming
from ChPT involves tree-level amplitudes and free propagators; there
is no possibility to compare with ChPT beyond leading order and thus
one is forced to fit the unknown low energy constants (LEC's) to data.

As it is the case in the purely mesonic sector, the off-shell scheme
generates a rich structure of unknown constants which allow for a good
description of the amplitudes.  Although the appearance of more
undetermined constants may appear a less predictive approach as, say,
putting a cut-off (one single parameter) in the divergent integrals as
it is done in Ref.~\cite{Na00}, it reflects the real state of the art
of our lack of knowledge on underlying QCD dynamics. The number of
adjustable LEC's should not be smaller than those allowed by symmetry;
this is the only way both to falsify all possible theories embodying
the same symmetry principles and to make wider the energy window which
is being described. Limiting such a rich structure allowed by CS 
results in a poor description of experimental data.

Before going further we would like also to say a word on the opposite
situation, i.e., the possibility of having more low energy parameters
than one actually needs. A possible redundancy of parameters is
obviously a undesirable situation. In standard ChPT the number of
LEC's is controlled to any order of the calculation by crossing and
unitarity. Moreover, the dependence of the observables on them is
strictly linear, so that it becomes possible to detect such a
redundant combination in case it occurs\footnote{A good example of
this is $\pi\pi$ scattering to two loops (~\cite{mksf95},~\cite{bc97})
where one gets, besides the four one loop parameters $\bar
l_{1,2,3,4}$, six new parameters but in redundant
combinations. Instead, it is customary to use the six $\bar
b_{1,2,3,4,5,6}$ independent combinations, which depend on the one
loop $\bar l$'s and the six new two loop parameters and contain mixed
orders. Such a situation also takes place in $\pi N$ scattering in
HBChPT at fourth order~\cite{fm00}}. In a unitarized approach, the only way to
avoid this parameter redundancy is to match the unitarized amplitude
to the standard ChPT amplitude. As we have already said, there is no
standard one loop ChPT calculation for the $S_{11}$ partial wave of
meson-baryon reaction with open channels to compare with.  An indirect
way to detect such a parameter redundancy might be through a fit to
experimental data if the errors in some parameters turn out to be very
large.

 We have considered four coupled channels: $\pi N$, $\eta N$, $K
\Lambda$, $K \Sigma$ and taken into account $SU(3)-$breaking symmetry
effects but neglected the considerably smaller isospin violation ones.
We have found that CS allows for a solution of the BSE which is
flexible enough to describe the elastic $\pi N $ phase-shifts, and the
$\pi^- p \to \eta n$ and $\pi^- p \to K^0 \Lambda$ cross section data
from threshold to CM energies well above the $N(1650)$ resonance.
Besides the rest of elastic and inelastic two-body reaction
channels\footnote{Each of the entries of the $4\times 4$ matrix
solution of the BSE is the $T-$scattering amplitude for a meson-baryon
reaction constructed out of the four considered channels: $\pi N \to
\pi N$, $\pi N \to \eta N$ , $\pi N \to K \Lambda$, $\pi N \to K
\Sigma$, $\eta N \to \eta N$, $\eta N \to K \Lambda$, $\eta N \to K
\Sigma$, $K\Lambda \to K \Lambda$, $K\Lambda \to K \Sigma$, $K\Sigma
\to K \Sigma$ and the reverse processes. } implicit in the adopted
formalism come as predictions of the model. The position of the
complex poles in the second Riemann sheet of the amplitudes determine
masses and widths of the $S_{11}-$ $N$(1535) and $-N$(1650)
resonances, which turn out to be in reasonable agreement with
experiment. Preliminary results were presented in \cite{bled2000}.

The paper is organized as follows: In Sect.~\ref{sec:thf} we present
the basic formalism used along the paper. We start with the chiral
Lagrangian relevant to our calculation, from which the lowest order
meson-baryon two particle irreducible matrix amplitude is
deduced. After presenting our notations for the coupled channel
kinematics we discuss the basic pertinent features of the BSE for
$s-$wave meson-baryon scattering. Using the amplitude from lowest
order ChPT as the potential, we solve and
renormalize the BSE in the spirit of an EFT.  In Sect.~\ref{sec:nr} we
present our numerical results, together with a detailed discussion on
the fitting procedure and Monte-Carlo propagation of inherited error
bars to all possible reaction channels. The quantum field theoretical
interpretation of resonances as unstable particles requires
determining their mass and width as poles in a unphysical Riemann
sheet. In our case there are 16 sheets which we discuss in some
detail, and we search for the most important pole singularities. Error
estimates are made in terms of the available experimental
uncertainties in the phase-shifts and amplitudes. Finally, in
Sect.~\ref{sec:concl} we present some conclusions and outlook for future work.

\newpage

\section{ Theoretical Framework} \label{sec:thf}

\subsection{Chiral Baryon-Meson Lagrangian}

At lowest order in the chiral expansion the chiral baryon meson
Lagrangian contains kinetic and mass baryon pieces and meson-baryon
interaction terms and is given by~\cite{Pich95}
\begin{eqnarray}
{\cal L}_1 = {\rm Tr} \left\{ \bar{B} \left( {\rm i}
\slashchar{\nabla} - M_B \right) B \right\} + 
\frac{1}{2} \, {\cal D} \, {\rm Tr} \left\{ 
\bar{B} \gamma^\mu \gamma_5 \left\{ u_\mu , B
\right\} \right\} + \frac{1}{2} \, {\cal F} \, {\rm Tr} \left\{ 
\bar{B} \gamma^\mu \gamma_5 [ u_\mu ,B] \right\}  \, ,
\label{LB1}
\end{eqnarray}
The meson kinetic and mass pieces and the baryon mass chiral
corrections are second order and read
\begin{eqnarray}
{\cal L}_2 &=& {f^2 \over 4} {\rm Tr} \left\{ u_\mu^\dagger u^\mu + 
(U^\dagger \chi + \chi^\dagger U ) \right\} \nonumber \\ &-&
b_0 {\rm Tr} ( \chi_+ ) {\rm Tr} (\bar B B) - b_1 {\rm Tr} ( \bar B
\chi_+ B ) - b_2 {\rm Tr} ( \bar B B \chi_+ )
\label{LB2}
\end{eqnarray}
where ``Tr'' stands for the trace in $SU(3)$. In addition,
\begin{eqnarray}
\nabla_{\mu} B &=& \partial_{\mu} B + \frac{1}{2}\, [ \, u^\dagger
\partial_{\mu} u + u \partial_{\mu} u^\dagger \, , \, B \, ] \,
,\nonumber \\
\label{LB1_exp}
U = u^2 &=& e^{ {\rm i} \sqrt{2} \Phi / f } \, , \qquad u_{\mu} = {\rm
i} u ^\dagger \partial_{\mu} U u^\dagger \, \nonumber \\ \chi_+ &=&
u^\dagger \chi u^\dagger + u \chi^\dagger u \, , \qquad \chi = 2 B_0
{\cal M}
\end{eqnarray} 
$M_B$ is the common mass of the baryon octect, due to spontaneous
chiral symmetry breaking for massless quarks. The $SU(3)$ coupling
constants which are determined by semileptonic decays of hyperons are
${\cal F} \sim 0.46$, 
${\cal D} \sim 0.79$ (${\cal F}+{\cal D} = g_{A} = 1.25$).  The constants
$B_0$ and $f$ (pion weak decay constant in the chiral limit) are not
determined by the symmetry. The current quark mass matrix is ${\cal
M}={\rm Diag}(m_u,m_d,m_s)$. The parameters $b_0$, $b_1$ and $b_2$ are
coupling constants with dimension of an inverse mass. The values of
$b_1$ and $b_2$ can be determined from baryon mass splittings, whereas
$b_0$ gives an overall contribution to the octect baryon mass
$M_B$. The $SU(3)$ matrices for the meson and the baryon octect are
written in terms of the meson and baryon spinor fields respectively
and are given by\footnote{For the purpose of our work we do not
consider any mixing between octect and singlet $SU(3)$
representations}
\begin{eqnarray}
	\Phi = \left( \matrix{ \frac{1}{\sqrt{2}} \pi^0 +
	\frac{1}{\sqrt{6}} \eta & \pi^+ & K^+  \cr  \pi^- & -
	\frac{1}{\sqrt{2}} \pi^0 + \frac{1}{\sqrt{6}} \eta & K^0  \cr 
	K^- & \bar{K}^0 & - \frac{2}{\sqrt{6}} \eta }
	\right) \, ,
\end{eqnarray}
and 
\begin{eqnarray}
	B =
	\left( \matrix{ 
	\frac{1}{\sqrt{2}} \Sigma^0 + \frac{1}{\sqrt{6}} \Lambda &
		\Sigma^+ & p \cr 
		\Sigma^- & - \frac{1}{\sqrt{2}} \Sigma^0 
		+ \frac{1}{\sqrt{6}} \Lambda & n \cr 
		\Xi^- & \Xi^0 & - \frac{2}{\sqrt{6}} \Lambda } 
	\right) \, .
\end{eqnarray}
respectively.  The $MB \to MB$ vertex obtained from the former
Lagrangian reads\footnote{We have omitted the pieces proportional to
the couplings ${\cal D}$ and ${\cal F}$ because they do not contribute
to $s-$wave. On the other hand, the lagrangian below does not lead to
a pure $s-$wave contribution and a further projection will be
required.}
\begin{equation}
\label{Lmbmb}
	{\cal L}_{MB \to MB} = \frac{\rm i}{4 f^2}{\rm Tr} \left\{
 \bar{B} \gamma^\mu \left[ \, [ \, \Phi \, , \, \partial_\mu \Phi \, ]
 \, , \, B \, \right] \, \right\} \, .
\end{equation}
Assuming isospin conservation, the scattering amplitude\footnote{We
use the convention, in symbolic notation, $-iT_{MB\to MB}
=+i{\cal L}_{MB\to MB}$.}  in the Dirac spinor basis, which
relation to the cross section is given in the next subsection, at
lowest order is given by
\begin{equation}
t_P^{(1)} (k,k') = {D \over f^2} ( \slashchar{k}+\slashchar{k}' )
\label{eq:lowest} 
\end{equation}
where $k$ and $k'$ are incoming and outgoing meson momenta and $D$ a
coupled-channel matrix. For strangeness $S=0$ and isospin $T=1/2$ the
coupled channel matrix $D$ reads\footnote{ There is a mistake in the
relative phases of Ref.~\cite{KSW95}. We thank A. Ramos for confirming
this point to us. We use the isospin phase convention of
Ref.~\cite{GP}: negative phases for the isospin states $ -|\pi^+
\rangle $ , $ -|\bar K_0 \rangle $ , $ -|\Sigma^+ \rangle $, $ 
-|\overline{\Xi^-}
\rangle $ , $ -|\overline{\Sigma^-} \rangle $ , $ -|\bar n \rangle $.  }

\begin{eqnarray} 
\nonumber \\ 
\hskip1.5cm 
\matrix{ \, \pi N \, &  \quad \eta N \, &  \, K
\Lambda \, & \, K \Sigma  & \qquad \qquad } \nonumber  
\\ 
D^{T=1/2}_{S=0} = {1\over 4} \left( \matrix{
 -2  &   0  &  -3/2  &   +1/2 \cr 
 0  &    0   & +3/2  &   +3/2 \cr
 -3/2  &   +3/2  &  0  &   0 \cr
 +1/2  &   +3/2  &  0  &  -2  
} \right)            
\qquad \matrix{   \pi N \cr  \eta N \cr   K \Lambda \cr  K \Sigma } 
\label{d-matrix} 
\end{eqnarray}

While amplitudes follow the chiral symmetry breaking pattern from the
effective Lagrangian to a good approximation, it is well known that
physical mass splittings have an important influence when calculating
the reaction phase space. Indeed, the correct location of reaction
thresholds requires taking physical masses for the corresponding
reaction channels. We have taken into account this effect in our
numerical calculation. Besides, chiral corrections to the amplitudes
also incorporate explicit CS breaking effects in addition to those
already present in the lagrangians above. In practice, we use
different numerical values for $f_\pi$ , $f_K$ and $f_\eta$, as it is
discussed in Sect.~\ref{sec:nr}. This can be easily accomplished through
the prescription 
\begin{eqnarray}
D/f^2 \to \hat f^{-1} D \hat f^{-1} \quad , \qquad \hat f \equiv {\rm
Diag} \left(f_\pi, f_\eta , f_K , f_K  \right)
\label{eq:f-presc}
\end{eqnarray}  
For simplicity and a more clear book-keeping of chiral order
dependences we will use the $D/f^2$ notation throughout the paper,
meaning Eq.~(\ref{eq:f-presc}) in practice. 

\subsection{Scattering Amplitude and Kinematics } 

The coupled channel scattering amplitude for the baryon-meson process
in the isospin channel, $T=\frac12$
\begin{eqnarray} 
B( M_A , P-k,s_A ) + M( m_A , k ) \to B( M_B , P-k', s_B) + M(
m_B , k' )
\end{eqnarray}
with baryon (meson) masses $M_A$ and $M_B$ ( $m_A$ and $m_B$ ) and
 spin indices (helicity, covariant spin, etc...) $s_A, s_B$,
 is given by
\begin{equation} 
T_P \left [ B \{k',s_B\}  \leftarrow 
 A \{k,s_A\}  \right  ] = \bar u_B ( P-k', s_B) t_P (k,k') u_A
 (P-k,s_A)\label{eq:deftpeque} 
\end{equation} 
Here, $u_A (P-k, s_A)$ and $u_B (P-k', s_B)$ are baryon Dirac
 spinors\footnote{We use the normalization ${\bar u }u = 2M$.} for the
 ingoing and outgoing baryons respectively, $P$ is the conserved
 total four momentum and $t_P (k,k') $ is a matrix in the Dirac and
 coupled channel spaces. On the mass shell and using the equations of
 motion for the free Dirac spinors $( \slashchar{P} - \slashchar{k} -
 M_A ) u_A (P-k)=0 $ and its transposed $ {\bar u}_A (P-k) (
 \slashchar{P} - \slashchar{k} - M_A ) =0 $ the parity and Lorentz
 invariant amplitude $t_P$ can be written as:
\begin{equation}
t_P (k,k') |_{\rm on-shell} = t_1 (s,t) \P + t_2 (s,t) 
\label{eq:amp-dirac}
\end{equation} 
with $s= P^2 = \P^2 $, $t = (k-k')^2$ and $t_1$ and $t_2$ matrices in
the coupled channel space. The normalization of the amplitude $T_P$ is
 determined by its relation to the CM differential
 cross section, and it is given above threshold, $s > {\rm Max} \{
 (M_A+m_A)^2, (M_B+m_B)^2 \}$,  by 
\begin{eqnarray}
\frac{d\sigma}{d\Omega} \left [B \{k_B,s_B\} \leftarrow A \{k_A,s_A\}
 \right ] &=& \frac{1}{64\pi^2 s} \frac{|\vec{k}_B|}{|\vec{k}_A|}
 \left |\, T_P \left [ B \{k_B,s_B\} \leftarrow A \{k_A,s_A\} \right ] \,
 \right |^2
\end{eqnarray}
Rotational, parity and time reversal invariances ensure
for the on shell particles
\begin{eqnarray}
T_P \left [ \{k_B,s_B\}  \leftarrow \{k_A,s_A\}  \right ]=
 -8\pi\sqrt{s} \sqrt{\frac{|\vec{k}_A|}{|\vec{k}_B|}}\left \{ {\cal
 A}(s,\theta )\delta_{s_A s_B } + {\rm i}\, {\cal
 B}(s,\theta ) \left ( \hat{n} \cdot \vec {\sigma} \right
 )_{s_B s_A } \right \}
\end{eqnarray}
${\cal A}$ and ${\cal B}$ are matrices in the coupled channel space, 
$\theta$  the CM angle between the initial and final meson three
momenta and $\hat{n}$  a unit three-vector orthogonal to
$\vec{k}_A$ and $\vec{k}_B$. Partial waves (matrices in the coupled
channel space), $f_{L}^J(s)$, 
are related  to ${\cal A}, {\cal B}$ by~\cite{EW88}  
\begin{eqnarray}
{\cal A}(s,\theta ) &=& \sum_L \left [ (L+1)\,\, f_{L}^{L+\frac12}(s) + 
L\,\, f_{L}^{L-\frac12}(s) \right ] P_L(\cos\theta) \nonumber\\
{\cal B}(s,\theta ) &=& -\sum_L \left [  f_{L}^{L+\frac12}(s) -  
\,\,f_{L}^{L-\frac12}(s) \right ] \frac{d P_L(\cos\theta)}{d\theta}
\end{eqnarray}
In terms of the matrices $t_1$ and $t_2$ defined in
Eq.~(\ref{eq:amp-dirac}), the $s-$wave coupled-channel matrix,
$f_0^\frac12 (s)$,  is  given by:
\begin{eqnarray}
\left[ f_0^\frac12 (s) \right]_{B \leftarrow A}&=& -\frac{1}{8\pi\sqrt s}
\sqrt{\frac{|\vec{k}_B|}{|\vec{k}_A|}} \sqrt{E_B + M_B}\sqrt{E_A +
M_A} \left [ \frac12 \int^1_{-1}
d\cos\theta 
\left ( \sqrt{ s} \,t_1(s,t) + t_2(s,t) \right )_{BA} \right ]
\label{eq:deff0}
\end{eqnarray}
where the CM three--momentum moduli read 
\begin{eqnarray}
|\vec{k}_i| &=& \frac{\lambda^\frac12 (s,M_i,m_i)}{ 2\sqrt {s}}\qquad i=A,B
\end{eqnarray}
with $\lambda(x,y,z) = x^2+y^2+z^2 -2xy-2xz-2yz$ and $E_{A,B}$ the
baryon CM energies. The phase of the matrix $T_P$ is such that the
relation between the diagonal elements ($A=B$) in the coupled channel
space of $f_0^\frac12 (s)$ and the inelasticities ($\eta$) and
phase-shifts ($\delta$) is the usual one,

\begin{eqnarray} 
\left [ f_0^\frac12 (s)\right]_{AA}  = {1\over 2 {\rm i} |\vec{k}_A| } 
\Big( \eta_A (s) e^{2
{\rm i} \delta_A (s)} - 1 \Big) 
\end{eqnarray}
Hence, the optical theorem reads, for $s \ge (M_A+m_A)^2$,
\begin{eqnarray}
\frac{4\pi}{|\vec{k}_A|} {\rm Im} \left [ f_0^\frac12 (s)\right]_{AA}  &=&
\sum_B \sigma_{B \leftarrow A} = 4\pi \sum_B \left |\left[  f_0^\frac12 (s) \right]_{BA}\right|^2 =  \sigma_{AA} + \frac{\pi}{|\vec{k}_A|^2}
\left(1-\eta_A^2\right) 
\end{eqnarray} 
where in the right hand side only open  channels contribute.

\subsection{ Bethe-Salpeter Equation }

The Bethe-Salpeter equation reads
 
\begin{equation} 
t_P ( k,k') = v_P ( k,k') + {\rm i} \int { d^4 q \over (2\pi)^4 }
t_P (q,k') \Delta(q) S(P-q) v_P (k,q) \label{eq:bse}
\end{equation}
where $t_P( k,k')$ is the scattering amplitude defined in
Eq.~(\ref{eq:deftpeque}), and $v_P(k,k')$ the two particle irreducible
Green's function (or {\it potential} ), and $ S(P-q)$ and $\Delta (q)
$ the baryon and meson exact propagators respectively. The above
equation turns out to be a matrix one, both in the coupled channel and
Dirac spaces. The resulting scattering amplitude $t_P( k,k')$ fulfills
the coupled channel unitarity condition
\begin{eqnarray}
t_P(k,k') - {\bar t}_P(k',k) &=& -{\rm i}(2\pi)^2 \int\frac{d^4
q}{(2\pi)^4} \nonumber \\ \times t_P(q,k')\, \delta^+ \left[ q^2-{\hat m}^2
\right] \left( \slashchar{P} - \slashchar{q} + {\hat M} \right) &
\delta^+ & \left[ (P-q)^2-{\hat M}^2 \right] {\bar t}_P(q,k)
\label{eq:off-uni}
\end{eqnarray} 
where $ {\bar t}_P(k,p) = \gamma_0 {t}_P^\dagger (k,p) \gamma_0 $ and
$ {t}_P^\dagger (k,p) $ stands for the total adjoint in the Dirac and
coupled channel spaces (including also the change $s + {\rm i}
\epsilon \to s - {\rm i} \epsilon $) and $\hat m$ and $\hat M$ the
meson and baryon (diagonal) mass matrices respectively. Finally, 
$\delta^+(p^2 - m^2) = \Theta(p^0) \delta(p^2-m^2)$, being $\Theta$
the Heaviside step function.

If the on shell amplitude depends only on the total
momentum\footnote{It is to say, the functions $t_1$ and $t_2$ in
Eq.~(\ref{eq:amp-dirac}) do not depend on the Mandelstam variable
$t$.} $P$, $t_P (k,k')|_{\rm on-shell} = t (\P) $, as it will be the
case below, the unitarity condition can be rewritten in a much simpler
and useful form as a discontinuity equation above the corresponding
physical thresholds
\begin{equation} 
{\rm Disc}[ t (\P)^{-1}] = - {\rm Disc} [ J(\P) ] \qquad {\rm with}
\qquad  {\rm Disc} [A(s)] \equiv
A(s+{\rm i}\epsilon) - A(s-{\rm i} \epsilon)  
\label{eq:iam}
\end{equation} 
where the quadratic and logarithmically divergent integral 
\begin{equation}
J(\P ) = {\rm i} \int { d^4 q \over (2\pi)^4 } {1\over q^2 -
\hat m^2}{1\over \slashchar{P}-\slashchar{q} - \hat M  } 
\end{equation}  
has been introduced and Cutkosky's rules used to evaluate its
 discontinuity. This integral is treated in detail in
 Appendix \ref{sec:app2}. As usual, we take the $ {\rm i} \epsilon $
 prescription $ \hat m^2 \to \hat m^2 - {\rm i} \epsilon $ and $ \hat
 M \to \hat M - {\rm i} \epsilon $ which we implicitly assume in the
 sequel.

\subsection{Solution of the BSE Equation at lowest order}

The BSE requires some input
potential and baryon and meson propagators to be solved. We proceed
here along the lines proposed in previous work~\cite{EJ99} and use a
chiral expansion to determine both potential and propagator. From the
chiral Lagrangian one gets at lowest order (Eq.~(\ref{eq:lowest})) 
\begin{equation}
v_P (k,k') = t_P^{(1)} (k,k') = {D \over f^2} ( \slashchar{k}+\slashchar{k}' ) 
\end{equation}
with $D$ the coupled-channel matrix, which was given in
Eq.~(\ref{d-matrix}). The propagators at lowest order are simply the
free ones,
\begin{equation}
\Delta(q) = {1\over q^2-\hat m^2 } \quad , \quad S(P-q)= {1\over 
\slashchar{P} - \slashchar{q} -\hat M } 
\end{equation}
which are diagonal in the coupled channel space. Once we have approximate
expressions for both the {\it potential} and the meson and baryon
propagators, we proceed to exactly solve the BSE. Second order Born
approximation to the solution of the BSE (i.e. approximating $t_P
(q,k')$ in the kernel of the equation by $v_P (q,k')$ ) suggests the
following form for the exact solution:
\begin{equation}
t_P (k,k') = a( \P) + \slashchar{k'} b_R ( \P)+ b_L ( \P) \slashchar{k}+ 
\slashchar{k'} c( \P) \slashchar{k}  
\label{eq:ansatz}
\end{equation} 
where $a$,$b_R$,$b_L$ and $c$ are Lorentz scalar matrices in the Dirac
and the coupled channel spaces. At this lowest order of the BSE
approach, these matrices only depend on $\P$, thus they turn out to be 
independent of the Mandelstam variable $t$.

\begin{figure}[tbp]
   \centering
   \footnotesize
   \epsfxsize = 13cm
   \epsfbox{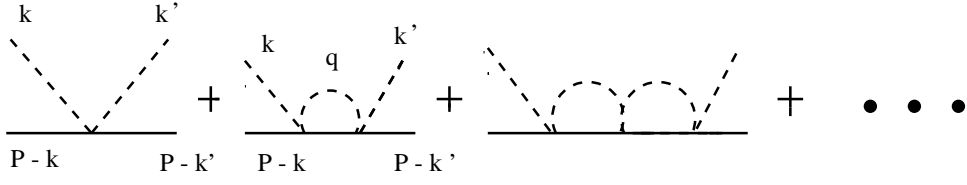} 
   \vspace*{1cm}

   \caption{ \footnotesize Diagrams summed by the Bethe Salpeter equation at
   lowest order. Kinematics defined in the main text.}  \label{fig_BS}
\end{figure}

Such an amplitude contains an infinite sum of diagrams, as shown in
Fig.~\ref{fig_BS}, but does not contain {\it all} possible one loop
dependences, for instance those coming from vertex renormalization. On
the mass shell we may set $\slashchar{k} \to \slashchar{P}-{\hat M} $ and
$\slashchar{k'} \to \slashchar{P}-\hat M $. Hence, thanks to the equations of
motion the on-shell amplitude becomes a function of the total momentum
$\P$ and reads
\begin{equation}
t (\P) = a + (\slashchar{P}-\hat M) \, b_R + b_L \, (
\slashchar{P}-\hat M) + (\slashchar{P}-\hat M) \, c \, (\slashchar{P}-\hat M)
\end{equation} 
where the explicit dependence on $\P$ of the $a,b_R,b_L,c$ matrix
functions has been suppressed for simplicity.

Plugging the ansatz for the off-shell amplitude, $t_P ( k,k') $, as
given in Eq.~(\ref{eq:ansatz}) into the BSE equation and after some
algebraic manipulations described in detail in the
Appendix~\ref{sec:app1}, we get for the inverse on-shell amplitude
\begin{eqnarray}
t (\P) ^{-1} &=& -J(\P) + {\Delta_{\hat m} \over \P - \hat M} + A^{-1}
\nonumber\\ && \nonumber\\ A &=& \frac{1}{f^2}  \{\P- \hat M,D \}_+ +
{1\over f^4} (\P- \hat M) D {\Delta_{\hat m} \over \P- \hat M } D (\P -\hat M)
 = \mu(s) + \nu(s) \P \nonumber \\ && \nonumber\\ \mu(s) &=& {1\over
f^4} \Big\{ - [D,\hat M] {\hat M \Delta_{\hat m}  \over s-\hat M^2 } 
[D,\hat M] -f^2 \{\hat M,D \}_+
-D\Delta_{\hat m} D \hat M + [D,\hat M]\Delta_{\hat m} D\Big\} \nonumber \\
&& \nonumber\\ 
\nu (s) &=& {1\over f^4} \Big \{ -[\hat M,D] \frac{\Delta_{\hat m}}{s-\hat M^2}
[\hat M,D] +
D\Delta_{\hat m} D + 2f^2 D \Big\}
\label{eq:t-1dirac}
\end{eqnarray} 
As we can see, the on-shell unitarity condition expressed in
Eq.(\ref{eq:iam}) is manifestly fulfilled. To proceed further we can
decompose the inverse on-shell amplitude in the form
\begin{equation}
t (\P) ^{-1}= K_1 (s) \P + K_2 (s)  
\label{eq:t-1}
\end{equation} 
where $K_1 (s)$ and  $ K_2 (s) $ are matrices in the coupled channel space. 
Straightforward calculation yields,   
\begin{eqnarray}
K_1 (s) &=& -{ s-\hat m^2+\hat M^2 \over 2 s} J_0 (s) + {\Delta_{\hat m} \over
s-\hat M^2} - \frac{\Delta_{\hat m \hat M}}{2s}
+ (\nu s - \mu \nu^{-1}\mu)^{-1} \nonumber  \\ && \nonumber \\
 K_2 (s) &=& -\hat M J_0 (s) +
 {\hat M \Delta_{\hat m}  \over s-\hat M^2} - (\nu s \mu^{-1}\nu
-\mu)^{-1}
\label{eq:defks}
\end{eqnarray} 
Thus, the amplitude can be written in the form  of Eq.~(\ref{eq:amp-dirac})
\begin{equation}
t (\P) = t_1 (s) \P + t_2 (s) \label{eq:tmatriz}
\end{equation} 
with 
\begin{eqnarray} 
t_1 &=& ( K_1 s - K_2 K_1^{-1} K_2 )^{-1} \nonumber\\
t_2 &=&  (K_2 - K_1 K_2^{-1} K_1 s )^{-1} 
\end{eqnarray}
As we already said, when the two particle irreducible
amplitude,$v_P$, and the meson and baryon propagators entering in the
BSE, the lowest ChPT order ones are taken, the matrices (in the coupled
channel spaces) $t_1$ and $t_2$ turn out to be independent of the
$t-$Mandelstam variable\footnote{Despite of that and because the Dirac
structure $\P$, the amplitude of Eq.~(\ref{eq:tmatriz}) not only
contains $s-$wave, but also a small $p-$wave.}. Hence the angle
integral in Eq.~(\ref{eq:deff0}) becomes trivial and apart from
kinematical factors, the relevant combination entering in the $s-$wave
scattering amplitude is
\begin{equation}
t(s) =  t_1(s) \sqrt{s} + t_2(s)  
= \left( K_1(s) \sqrt{s} + K_2(s) \right)^{-1} \label{eq:defts}
\end{equation} 
After some algebraic reshuffling the expression for the inverse
coupled channel matrix amplitude can conveniently be written as
\begin{eqnarray}
t(s)^{-1} &=& -{ (\sqrt{s}+\hat M)^2-\hat m^2  \over 2 \sqrt{s}} \hat
J_0 (s) + {\Delta_{\hat m} \over \sqrt{s}-\hat M} - \frac{\Delta_{\hat
m \hat M}}{2 \sqrt{s}} \nonumber \\ &+& \left[ \frac1{f^2} \left\{
\sqrt{s}-\hat M, D \right\}_+ + \frac1{f^4} \left( \sqrt{s}-\hat M 
\right) D \frac{\Delta_{\hat m}}{\sqrt{s}-\hat M} D \left(
\sqrt{s}-\hat M \right) \right]^{-1} 
\label{eq:s11-amp}
\end{eqnarray}
Finally, the $s-$wave coupled-channel matrix
amplitude $f_0^\frac12 (s)$ reads, 
\begin{eqnarray}
\left[ f_0^\frac12 (s) \right]_{BA} &=& -\frac{1}{8\pi\sqrt s}
\sqrt{\frac{|\vec{k}_B|}{|\vec{k}_A|}} \sqrt{E_B + M_B}\sqrt{E_A + M_A}
\Big[ t(s) \Big] _{BA}
\end{eqnarray}
At this point we have to renormalize the divergent integrals appearing
in the solution of the BSE. This issue has been carefully discussed in
the context of meson-meson scattering (see Sect. 3.4 of
Ref.~\cite{EJ99}) and applies equally well in the present context. In
summary, the result of that discussion amounts to write the
renormalized amplitudes (finite) in the same way as the divergent
amplitudes but with the previously divergent integrals taken as finite
renormalized constants. Ideally one would like to derive them from the
underlying QCD dynamics, but in practice it proves easier to fit them
to experiment.  This amounts to consider, besides the physical masses
and weak meson decay constants 12 fitting parameters that define three
diagonal matrices in the coupled channel space which appeared already
in the solution of the BSE given, e.g., in Eq.~(\ref{eq:t-1}). These
matrices are
\begin{eqnarray} 
J_0(s=(\hat m + \hat M)^2) &=& \left ( \matrix{ J_{\pi N}
& 0 & 0 & 0 \cr
0& J_{\eta N} & 0 & 0 \cr 
0& 0& J_{K \Lambda} & 0 \cr
0& 0& 0& J_{K \Sigma} } \right ) \nonumber \\ && \nonumber \\ 
 \Delta_{\hat M} &=& \left ( \matrix{  \Delta_{N, 1} 
& 0 & 0 & 0 \cr
0&  \Delta_{N, 2} & 0 & 0 \cr 
0& 0&  \Delta_{\Lambda} & 0 \cr
0& 0& 0&  \Delta_{\Sigma} } \right ) \nonumber \\ 
&& \nonumber \\ 
 \Delta_{\hat m} &=& \left ( \matrix{  \Delta_{\pi } 
& 0 & 0 & 0 \cr
0&  \Delta_{\eta } & 0 & 0 \cr 
0& 0&  \Delta_{K,1} & 0 \cr
0& 0& 0&  \Delta_{K,2} } \right ) 
\end{eqnarray} 
where we have denoted the meson-baryon low energy constants $J_0
(s=(m_i+M_j)^2 )$, $i=\pi,\eta,K,K$ and $j=N,N,\Lambda,\Sigma$ of
Eq.~(\ref{eq:J0mM}) as $J_{ij}$.  The matrix $\Delta_{\hat m , \hat
M}$ which appears in Eq.~(\ref{eq:t-1}) is determined by the matrices
$\Delta_{\hat m}$ and $\Delta_{\hat M}$ above, as it is defined in
Eq.~(\ref{eq:deltafit}). On the other hand, the $\Delta_{K,1}$ and the
$\Delta_{K,2}$ matrix elements of the matrix $\Delta_{\hat m}$ and the
$\Delta_{N,1}$ and the $\Delta_{N,2}$ matrix elements of the matrix
$\Delta_{\hat M}$ have been taken in general to be different. This is
because, though their formal expression as divergent integrals in
Eqs.~(\ref{eq:defdelta}) and~(\ref{eq:deltafit}) are the same, after
doing the needed renormalization there is no reason why the finite
parts in these two channels should coincide\footnote{This point is
clearly exemplified in the $\pi\pi$ BSE treatment, see Eq.~(A.15) in
Ref.~\cite{EJ99}, where constants which stem from the same divergent
integrals, after renormalization become in fact different functions of
the $SU(2)$ low energy constants $\bar l$'s.}.

\section{Numerical results} \label{sec:nr}

Throughout the paper we will use the following numerical values for 
masses and weak decay constants of pseudoescalar mesons (all in MeV), 
\begin{eqnarray}
m_\pi = 139.57 \qquad m_\eta &=& 547.45 \qquad m_K = 497.67 \nonumber
\\ M_N = \cases{ M_p= 938.27 \cr M_n = 939.57 } \qquad M_\Lambda &=&
1115.68 \qquad M_\Sigma = 1192.55 \nonumber \\ f_\pi = 93.2 \qquad
f_\eta &=& f_K = 1.3 f_\pi
\label{eq:numval}
\end{eqnarray} 
where for the channel 11 the proton mass is used, because the data
have been obtained from the $\pi^- p $ scattering, and for the channel
22 we take the neutron mass, because the available data come from
$\pi^- p \to \eta n $. In this way we ensure the exact and physical
position of the thresholds. This proves important due to the proximity
of the $N(1535)$ to the $\eta n $ threshold.

\subsection{ Fitting procedure} 

We perform a $\chi^2-$fit with 12 free parameters considering the following
experimental data and conditions:
\begin{itemize}
\item 
\underline{ $S_{11}$ $\pi N $ elastic phase shifts and
inelasticities\cite{AS95}, $1077.84 \, {\rm MeV} \le \sqrt{s} \le
1946.52 \, {\rm MeV}$}: In this CM energy region, there are a total
number of 281 phase shifts and inelasticity data points . Though we
have considered four coupled channels, the three-body $\pi \pi N$
channel is not explicitly considered.  This omission influences both
the phase shifts and the inelasticities and we will assume here that
the effect is much more important for the inelasticities than for the
phase shifts.  Thus, we have fitted the phase-shifts while
inelasticities have been considered only to impose some constraints on
the fit. While $\eta > 0.99$, we have considered that the $\pi\pi N$
channel is essentially closed. In the data, this is the case for CM
energies below $\sqrt{s}=1406.4 \, {\rm MeV}$. In this energy region,
we have assigned to the phase shifts a $3\%$ relative error added in
quadrature with a systematic $1^o$ absolute error, in the spirit of
Ref.~\cite{BKM97a}. In this way, we are assuming that any $\pi\pi N$
subthreshold effects are effectively incorporated in the systematic
error mentioned above. At higher energies, inelasticities are smaller
than 0.99 and we provide the phase shifts with a systematic $15^o$
absolute error added in quadrature with a $3\%$ relative error. The
reason for this big systematic error is to account for the explicit
omission of the, now open and likely important, three-body channel.

Despite of being able to account for a part of the inelasticities
($\eta N$, $ K \Lambda $ and $ K \Sigma $ channels ) and because of
the explicit omission of the $\pi\pi N$ channel, inelasticities have
not been fitted. Nevertheless, some constraints are imposed in order
to prevent the occurrence of smaller inelasticities than the
experimental ones.  On a quantitative level this means the
following. Firstly, we provide the inelasticities with a $3\%$
relative error added in quadrature with $0.01$ absolute error. In the
second step within the $\chi^2-$fit procedure and given a set of
parameters, we compute the theoretical inelasticity for each
$\sqrt{s}$. If it turns out that for one CM energy the inelasticity is
smaller than the experimental value, taking into account the provided
errors, we strongly disfavor this set of parameters by decreasing the
total error on the inelasticity for this CM energy by an order of
magnitude when calculating its contribution to the total
$\chi^2$. Besides, those energies for which the theoretical
inelasticities fall above the experimental ones are set to 
contribute zero to the
total $\chi^2$. In this way, we do not force the fit to pass through
the experimental inelasticities at all, but avoid the unphysical
scenario where $\sum_{i=\eta N, K\Lambda , K \Sigma} \, \,
\sigma_i^{\rm theoretical} > \sigma_{\rm inel}^{\rm experimental} $.

\item \underline{Total $\pi^- p \to \eta n $ cross section
\cite{Ba88}, $1488.4 \, {\rm MeV} \le \sqrt{s} \le 1563.8 \, {\rm
MeV}$}

We fit the region close to the $\eta n $ threshold and in terms of the
commonly used $q_{\rm LAB}$ (incoming pion momentum in the laboratory
(LAB) system), the above range corresponds to $687 \, {\rm MeV}
\lesssim q_{\rm LAB} \lesssim 812 \, {\rm MeV}$. There is a total
number of 11 data points. The experimental uncertainties are provided
in Ref.~\cite{Ba88}.  In addition, the experimental cross section has
the contribution not only of the $s-$wave, object of this work, but
also of the rest of higher partial waves. Next to threshold the
$s-$wave is the dominant contribution and the higher energy cut
(1563.8 MeV) determines the region up to where it is still a good
approximation to the total cross section. For higher energies the
$p-$wave does play an important role and cannot be neglected,
Ref.~\cite{Caro00}.

We have neglected any possible effect stemming from the $\pi \pi N$
intermediate state in this inelastic channel, as it has been also
assumed previously in Refs.~\cite{KSW95} and~\cite{Na00}.

\item \underline{ Total $\pi^- p \to K^0 \Lambda $ cross section
\cite{Ba88}, $1617.5 \, {\rm MeV} \le \sqrt{s} \le 1724.8 \, {\rm
MeV}$}.

We fit the region close to the $K^0 \Lambda $ threshold, $904 \, {\rm
MeV} \lesssim q_{\rm LAB} \lesssim 1097 \, {\rm MeV}$ with the
experimental error bars provided in Ref.~\cite{Ba88}. There is a total
number of 45 data points and the remarks concerning both the
contribution of the $\pi\pi N$ channel and $p-$ wave effects of the
previous item apply also here. 
 
\end{itemize} 

Note that we have not fitted the $\pi^- p \to K^0 \Sigma^0 $ total
cross section because of the likely sizable isospin $3/2$
contribution.

\begin{figure}[]

\begin{center}                                                               
\epsfysize = 650pt
\makebox[0cm]{\epsfbox{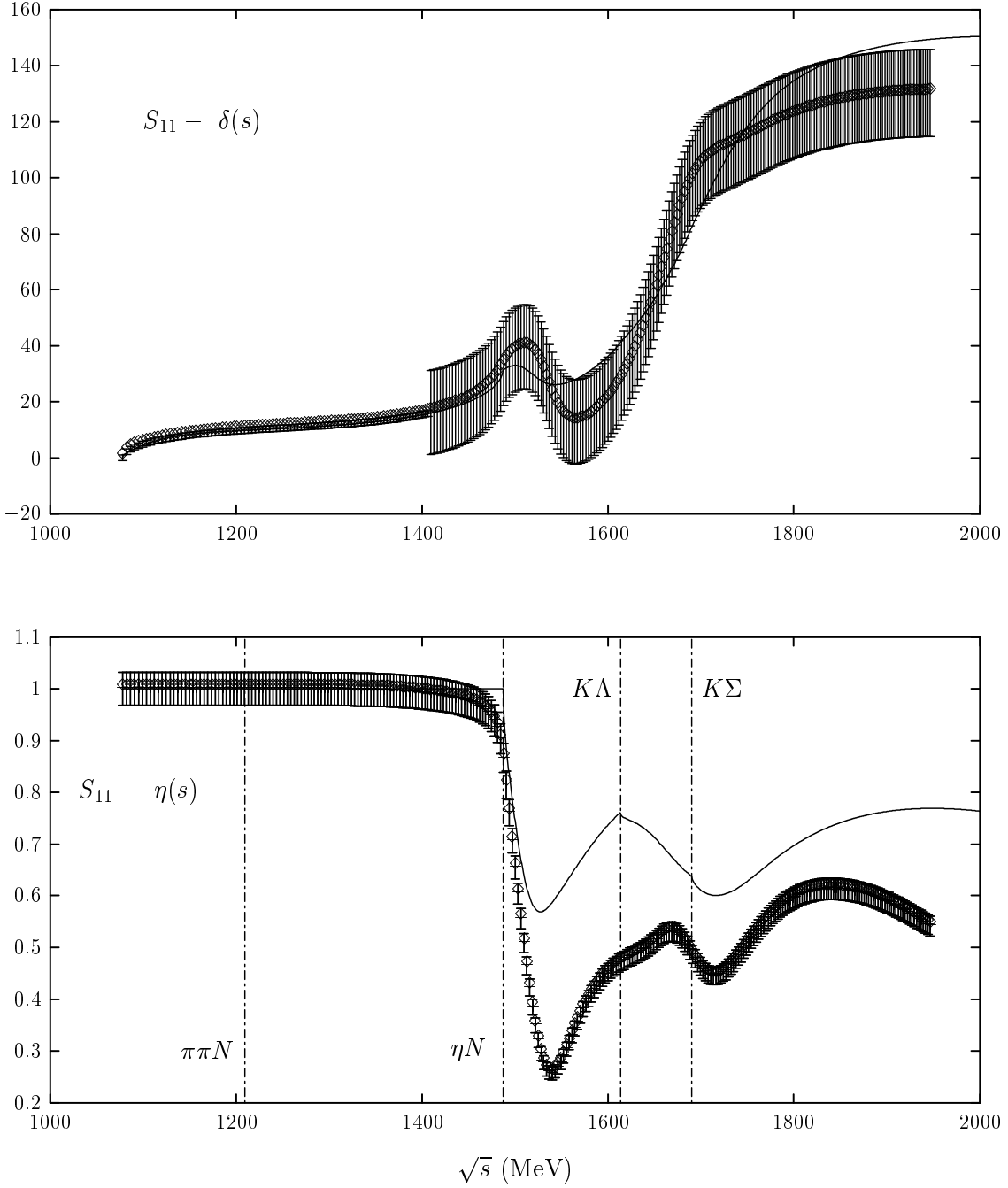}}
\end{center}
\vspace{-3.5cm}
\caption[pepe]{\footnotesize $S_{11}-$elastic $\pi N $ phase--shifts
and inelasticities as a function of CM energy $\protect{\sqrt{s}}$.
Data from Ref.~\cite{AS95}. Solid lines stand for the lowest--order
BSE results with parameters given in Appendix~\ref{sec:app-stat}.
Dotted-dashed vertical lines in the bottom plot indicate the energies 
for which new channels are opened.} 
\label{fig:pha_ine}
\end{figure}
\begin{figure}[]
\begin{center}                                                                
\epsfysize = 600pt
\makebox[0cm]{\epsfbox{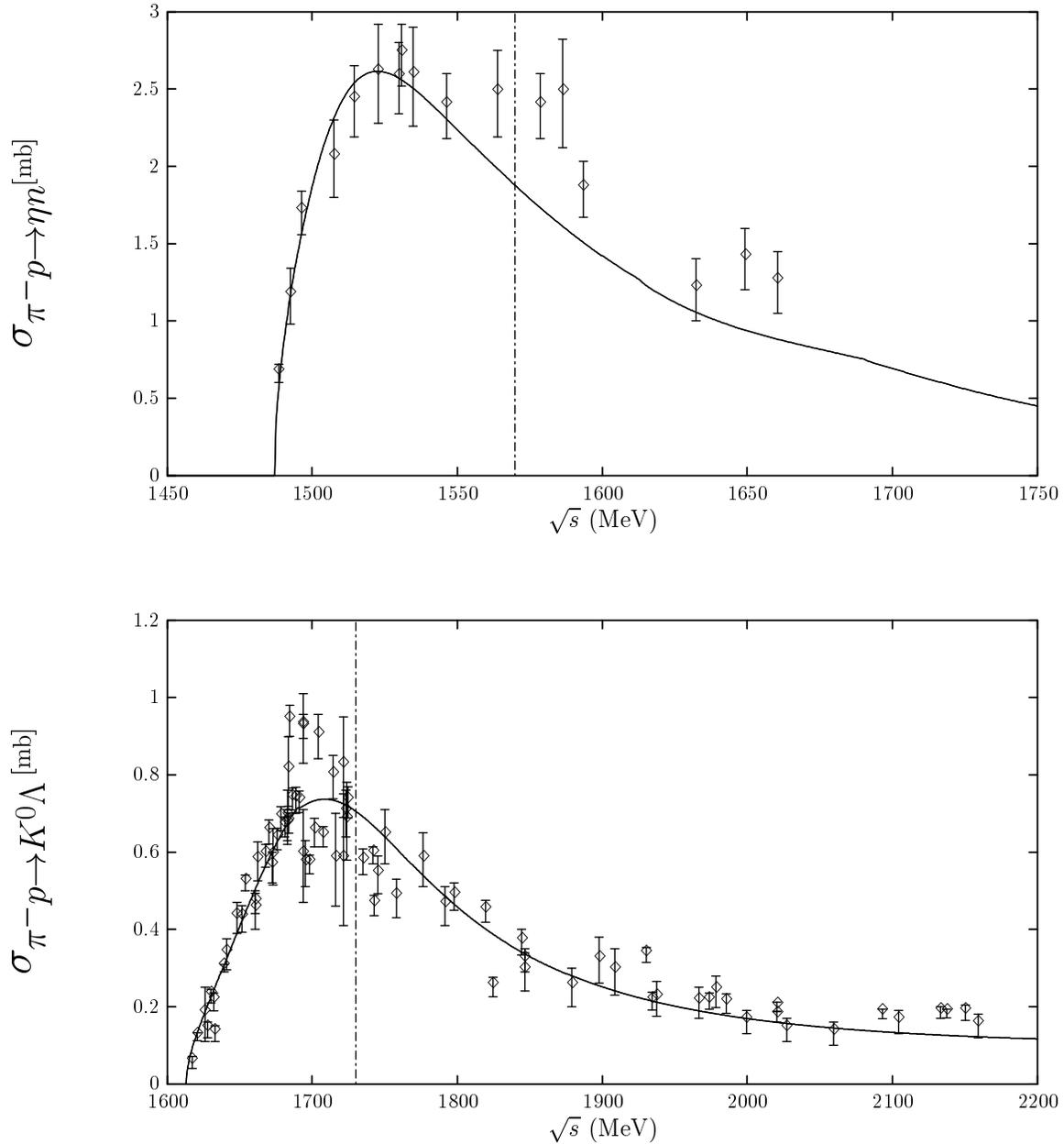}}
\end{center}
\vspace{-3.5cm}
\caption[pepe]{\footnotesize Total $\pi^- p \to \eta n$ and $\pi^- p
\to K^0 \Lambda $ cross sections as a function of the CM energy. Data
 from Ref.~\cite{Ba88}. Solid lines stand for the lowest--order
BSE $S_{11}-$results with parameters given in
Appendix~\ref{sec:app-stat}. Data for energies above the vertical
dotted-dashed lines have not been included in the fit.}
\label{fig:solo_sec}
\end{figure}

\subsection{Results of the best $\chi^2-$fit}

The best fit parameters are compiled in
Appendix~\ref{sec:app-stat}. The errors on the parameters turn out to
be fairly small.  They are purely statistical and have been obtained
from the $68\%$ confidence level on the best-fit parameter
12-dimensional distributions.  We generate these parameter
distributions out of $N=10^4$ samples. Each of the samples is obtained
from a $\chi^2-$fit to a synthetic set of data--points which are
obtained from the original one by a Gaussian sampling procedure,
i.e. the total number of fits is $N=10^4$.  We use the so obtained
distributions to evaluate the correlation matrix between the
parameters. The correlation matrix is also given in
Appendix~\ref{sec:app-stat}. Besides, quantities coming from a
$\chi^2-$fit, are Gauss distributed in the limit of small errors, as
it seems to be the case here, therefore the correlation matrix
determines the parameter distributions. 

Systematic errors on the parameters are not included and they are difficult to 
estimate. We can not completely discard that they might be
sizeable. This situation reflects the present status of the art in
unitarized calculations. 

In Figs.~\ref{fig:pha_ine} and~\ref{fig:solo_sec} we show the results
of our approach, with parameters given in Appendix~\ref{sec:app-stat},
for those quantities which have been fitted to. In
Fig.~\ref{fig:solo_sec} only the data for energies below the vertical
dotted-dashed lines have been included in the fit. The overall
description is remarkably good and that gives us some confidence on
the used non-perturbative resummation procedure based on the BSE. For
the elastic $\pi N \to \pi N$ scattering length we get
\begin{equation}
a^\frac12_0 \equiv 
\left [ f^\frac12_0(s=(m_\pi+M_N)^2)\right]_{\pi N \leftarrow \pi N} = 
0.179 \pm 0.004 \quad {\rm fm}
\label{eq:apin} 
\end{equation}
where the error is statistical and it has been obtained from those in
the best fit parameters (Eq.~(\ref{eq:lecs})), taking into account the
existing statistical correlations, through a Monte--Carlo
simulation. This value should be compared both to the recent
experimental one $0.252 \pm 0.006$ fm of Ref.~\cite{Sch99} and to the
HBChPT result to third order of Ref.~\cite{ej00c} $0.19 \pm 0.05$
fm. The agreement between our coupled channel unitarized scattering
length with that from NNLO-HBChPT is satisfactory from a theoretical
viewpoint since in both cases the same set of low energy
data~\cite{AS95} have been used.  The discrepancy of our number with
the experimental one of Ref.~\cite{Sch99} possibly points toward a too
conservative error assignment of the low energy phase-shifts in
Ref.~\cite{AS95}.

In principle, the LEC's of Eq.~(\ref{eq:lecs}) determine, or viceversa
they can be determined from, the next-to-leading order results of
ChPT, as it is explicitly shown in Ref.~\cite{EJ99}, for the case of
elastic $\pi\pi-$scattering. However, the perturbative calculation is
not available and it only exists next-to-leading results for $\pi N$,
with no coupled channels~\cite{BKM97a,fm00,Mo98,Fettes98}, in
the framework of Heavy Baryon Chiral Perturbation Theory (HBChPT). 
In Appendix~\ref{sec:matching} we discuss this point in more detail.

\subsection{ Predictions for other processes} 

\begin{figure}[]
\begin{center}                                                               
\leavevmode
\epsfysize=650pt
\makebox[0cm]{\epsfbox{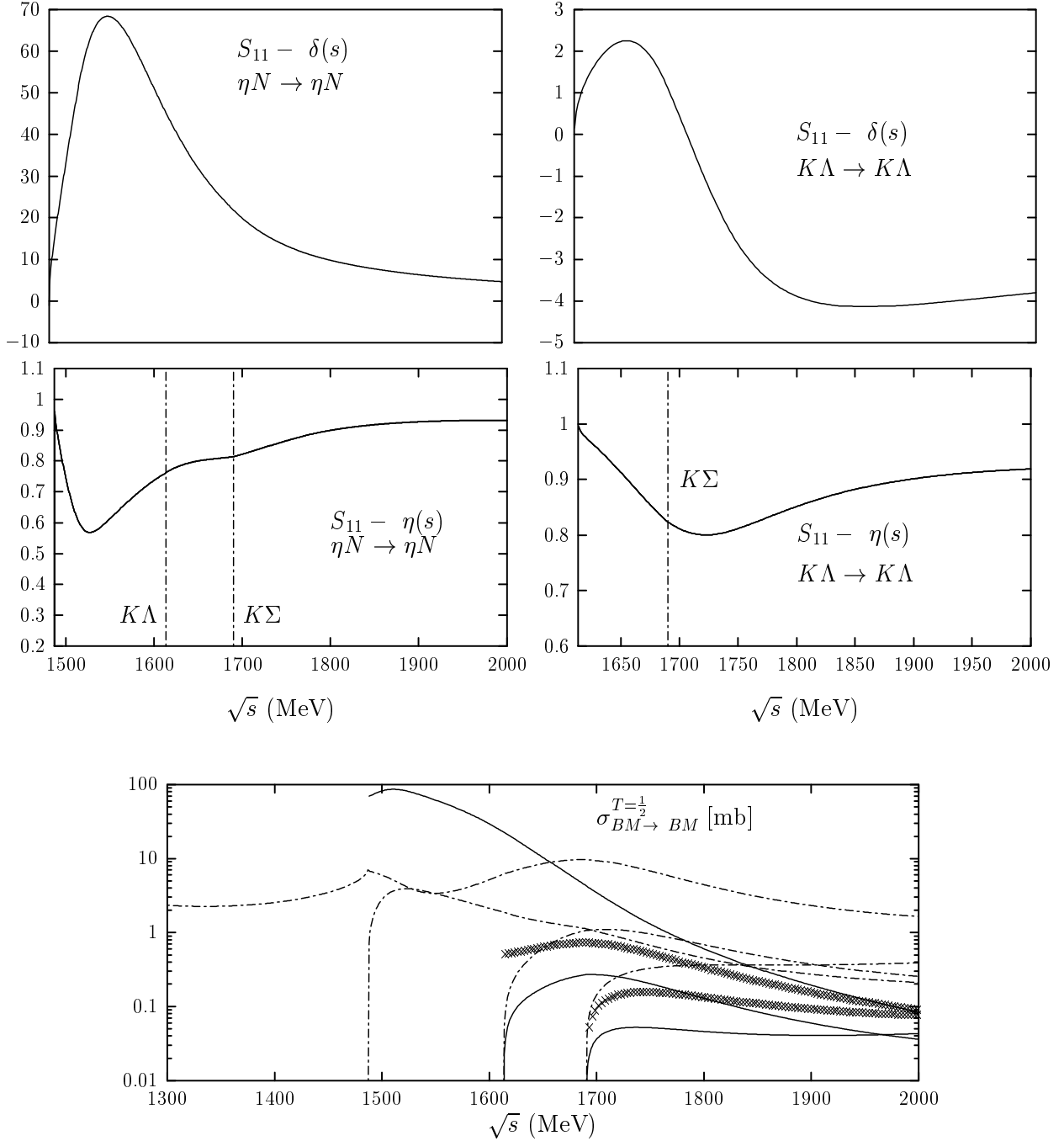}}
\end{center}
\vspace{-3.4cm}
\caption[pepe]{\footnotesize Top pannel: $s-$wave $T=1/2$ phase shifts
(in degrees) for elastic $\eta N \to \eta N$ (left pannel) and $ K
\Lambda \to K \Lambda $ (right pannel) processes as functions of the
CM energy. Middle pannel: same as before but for
inelasticities. Vertical lines indicate the opening of reaction
thresholds. Bottom pannel: $s-$wave $T=1/2$ meson-baryon cross
sections ($\pi N \to \pi N , \eta N , K \Lambda , K \Sigma$; $\eta N
\to \eta N , K \Lambda , K \Sigma$; $K \Lambda \to K \Lambda, K
\Sigma$)  in mbarns as functions of the CM energy. Dashed lines
indicate $\pi N$ initial state. Solid lines indicate $\eta N$ initial
state. Crosses indicate $K \Lambda$ initial state. All the lines start
at the relevant final state threshold (with the exception of the
elastic $\pi N \to \pi N$ reaction).}
\label{fig:2233}
\end{figure}

In Fig.~\ref{fig:2233} we show some of our predictions for phase
shifts, inelasticities and $s-$wave $T=1/2$ partial cross sections for
some other channels. For most of them there are no data. The  $\eta N \to
\eta N$ elastic phase shifts (top left pannel) present a steep raise
close to the threshold going up to $70^o$, which corresponds to a
typical low energy resonance behaviour triggered by the $N(1535)$
resonance (see next section) . Accordingly, the corresponding partial
cross section (bottom pannel) takes a unnaturally large value as
compared to other elastic and transition cross sections. This
is in contrast to any expectation based in the Born approximation,
since the corresponding potential in this channel vanishes (see
Eq.(~\ref{d-matrix})). 
\vspace{-2cm}
The effect of the $N(1535)$ can also be seen at the figure (bottom
pannel) by the maximum in $\pi N \to \eta N $ cross section and the
cusp effect in the $\pi N \to \pi N$ partial cross section.  On the
other hand, $K \Lambda \to K \Lambda $ phase shifts (top right pannel)
turn out to be extremely small. The effect of the $N(1650)$ can also
be seen at the cross sections, particularly in the $\pi N \to \pi N$
partial cross section, though the effect is less pronounced than in
the $N(1535)$ case.  We have not plotted the elastic $K \Sigma \to K
\Sigma $ cross section since the physical process involves also the
isospin $T=3/2$ channel, not considered in this work.

Our estimates for $\eta N$ and $K \Lambda$ scattering lengths (defined
for the elastic channels similarly as in Eq.~(\ref{eq:apin}) are
\begin{eqnarray} 
a_{\eta N}    &=& 0.772 (5) \,\, + \, {\rm i} \, 0.217 (3) \, {\rm
fm}\nonumber  \\ 
a_{K \Lambda} &=& 0.0547 (5) \, + \, {\rm i} \, 0.032 (4) \, {\rm fm} 
\end{eqnarray} 
respectively. The scattering length $a_{\eta N}$ compares reasonably
well with the one obtained in Ref.~\cite{KSW95}, $a_{\eta N} = 0.68 +
{\rm i} 0.24 $ fm.

\subsection{ Second Riemann sheet: poles and resonances.}

\begin{figure}[tbp]
   \centering
   \footnotesize
   \epsfxsize = 15cm
   \epsfbox{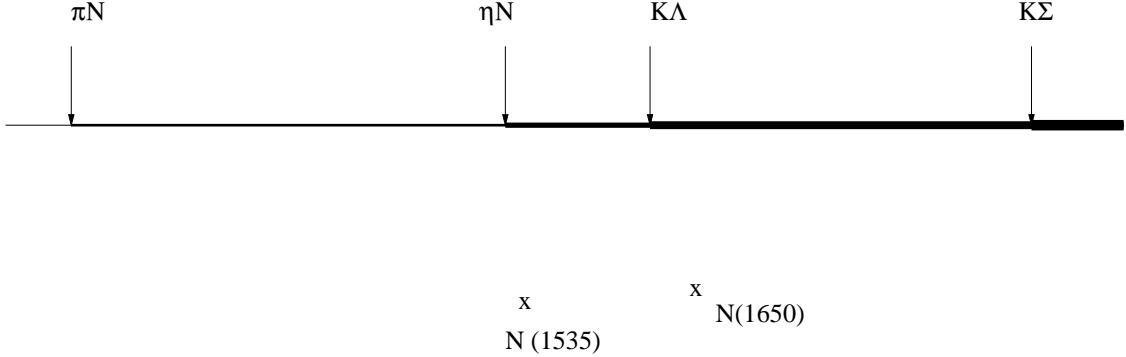} 
   \vspace*{1cm}
   \caption{ \footnotesize Location of reaction thresholds and resonances in
    the complex CM energy plane. The corresponding unitarity cuts have
    increasing thickness for increasing energy }
    \label{fig:thresholds}
\end{figure}

In this section we are interested in describing masses and widths of
the $S_{11}-$ resonances. An illustrative picture of the complex CM
energy plane with the singularities from the Particle Data
Book~\cite{PDG98} is presented in Fig.~\ref{fig:thresholds}. For a
more distinctive characterization of the resonances one has to look
for poles in the complex $s-$ plane. 

Since causality imposes
the absence of poles in the physical sheet~\cite{Ma58}, one should
search for complex poles in unphysical ones. Among all of them, those
{\it closest} to the physical sheet are the most relevant ones. For
the sake of clarity, we will devote some space here to explain, in a
quantitative manner, the meaning of ``close'' in this context. We look
for poles in the coupled channel matrix amplitude $t(s)$ defined in
Eq.~(\ref{eq:defts}). We have only examined the entry 11,
$\pi N \to \pi N$, of that matrix. The position of the complex poles,
as long as they are produced for physical resonances, should be
independent of the particular channel. However, residues at the pole do
depend on the examined channel, because they determine the coupling of
each of the channels to the resonances. This interesting point will be
discussed elsewhere~\cite{ej01b}.

To begin with, let us assume a situation where the coupled channel
 formalism is not needed, i.e. an artificial situation where only the
 11 element of the first column and row of the $D-$matrix in
 Eq.~(\ref{d-matrix}) is non-vanishing. In such a case, elastic
 unitarity  requires only a unique finite branch point at $s= s_+=
 (M_N + m_\pi)^2$ and a cut along the line $[s_+,+\infty[$. The scattering
 amplitude in the unphysical second Riemann sheet ($t_{II}(s)$)
 is simply obtained by analytical continuation of the amplitude in the 
 physical first  Riemann sheet ($t(s) \equiv t_I(s)$) across the
 unitarity cut,
 and therefore the following relation for inverse amplitudes should hold
($s$ real and above $s_+$)
\begin{eqnarray} t_{II}^{-1} (s+{\rm i}\epsilon ) &=& t_I^{-1} (s-{\rm
 i}\epsilon ) \label{eq:continuidad}
\end{eqnarray}

\vspace{1cm} 

The unitarity condition for the inverse
amplitude, deduced from Eqs.~(\ref{eq:defts}),(\ref{eq:defks})
and~(\ref{eq:disj0}) reads, 
\begin{eqnarray}
{\rm Disc} \, [t^{-1} (s)] \equiv t_I^{-1} (s+{\rm i}\epsilon
)-t_I^{-1} (s-{\rm i}\epsilon ) = 2 {\rm i} \rho(s)
\qquad s > s_+
\end{eqnarray} 
with $s \in \reales$, where the phase space function
\begin{eqnarray}
\rho(s) = {\lambda^{1/2} (s,m_\pi,M_N) \over 16 \pi s } \times 
{ ( \sqrt{s}+M_N)^2 -m_\pi^2  \over 2 \sqrt{s} }
\end{eqnarray} 
has been introduced, understanding that $\rho(s)$ is a function of the
real variable $s$.  Then, analytically continuing the phase space
function to all complex plane, the unitarity condition reads
\be
 t_I^{-1} (s+{\rm i}\epsilon)-t_I^{-1} (s-{\rm i}\epsilon ) 
= 2 {\rm i} \rho(s+{\rm i}\epsilon) \qquad s_+ < s \in \reales \label{eq:39}
\ee
where the cuts for $\lambda^\frac12 (z,m_\pi^2,M_N^2)$ go along the
real axis for $-\infty < s < s_- $ and $ s_+ < s < \infty $. The
function is chosen to be real and positive on the upper lip of the
second cut, $ s_+ < s < \infty $ and hence it satisfies:
\begin{eqnarray}
\lambda^\frac12 (s+{\rm i}
\epsilon ,m_\pi^2,M_N^2) = -\lambda^\frac12 (s-{\rm i} \epsilon
,m_\pi^2,M_N^2) = |\lambda^\frac12 (s,m_\pi^2,M_N^2)| \qquad  s_+
< s  \in \reales
\end{eqnarray}
 Besides, the cut for the $\sqrt{z}$ function, also appearing in
$\rho(z)$, is taken along the line $]-\infty,0]$ and the multivalued
function is taken to be positive for real and positive values of
$z$. Now using Eqs.~(\ref{eq:continuidad}) and~(\ref{eq:39}) one
finds the amplitude in the second Riemann sheet,

\be 
t_{II}^{-1} (z) = t_I^{-1} (z)-2 {\rm i} \rho(z) \qquad z \in \comp 
\ee
On the other hand, 
\begin{eqnarray} 
t_{II}^{-1} (s - {\rm i}\epsilon)& = & t_I^{-1} (s - {\rm i}\epsilon)
-2 {\rm i} \rho(s - {\rm i}\epsilon) = t_I^{-1} (s + {\rm i}\epsilon)
-2 {\rm i} \rho(s + {\rm i}\epsilon) -2 {\rm i} \rho(s - {\rm
i}\epsilon) \nonumber \\ &&  \nonumber \\
& = & t_I^{-1} (s + {\rm i}\epsilon) \qquad s_+ < s \in \reales
\end{eqnarray}
which means that there are only two Riemann sheets  linked to the
unitarity cut. The analytical structure, concerning the unitarity cut,
of the inverse amplitude  is determined by  the function
\begin{eqnarray}
-\bar J_0(s) &\times& (\frac{s-m_\pi^2+ M_N^2}{2\sqrt s} + M_N)
\end{eqnarray}
as it is deduced from Eqs.~(\ref{eq:defts}), (\ref{eq:defks})
and~(\ref{eq:J0mM}). Because of the choice of cuts for the
multivalued function $\sqrt{z}$ above, the function $L(s)$ introduced
in Eq.~(\ref{eq:defls}) determines the analytical structure of
$t(s)$. Thus, the two Riemann sheets of $t(s)$ related to
the unitarity cut, are obtained from the values $n=0$ and 1 in
Eq.~(\ref{eq:L(z,n)}) for $L(z,n)$.

In the general case of multiple thresholds, as it is the case in this
work, the above conclusions hold for any of the finite branch points,
located at the physical thresholds, and unitarity related cuts going
along the lines [branch point, $\infty$[. 

\begin{figure}[tbp]
   \centering \footnotesize \epsfxsize = 13cm \epsfbox{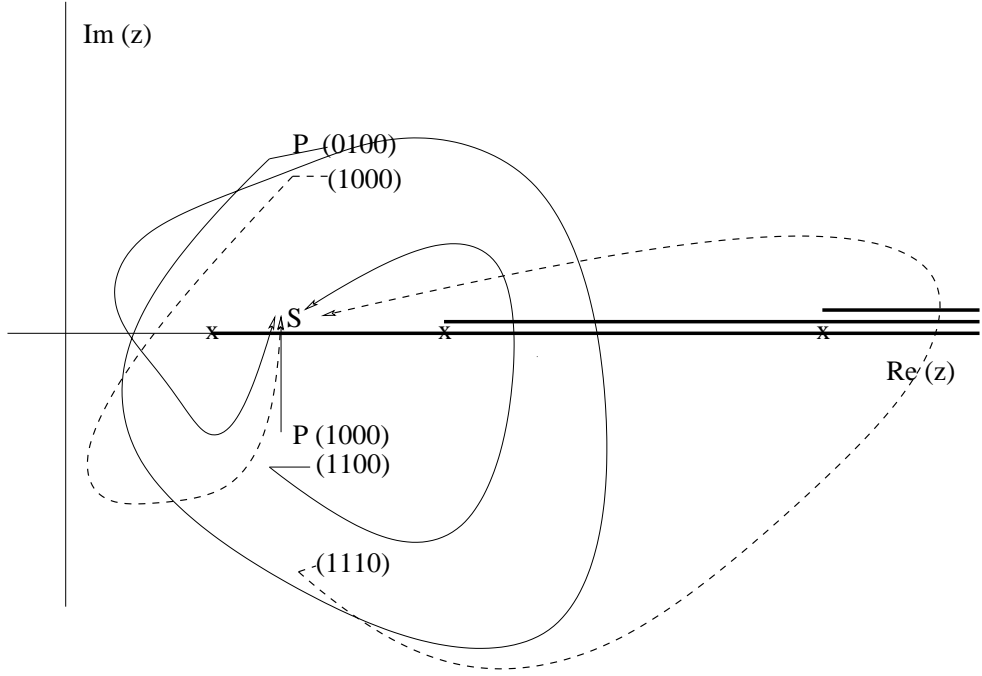}
   \vspace*{1cm} \caption{ \footnotesize Different paths, in the
   $s-$complex plane, showing how to reach the point $S$, located in the
   physical scattering region in the interval $ \left[(m_\pi+M_N )^2,
   (m_\eta + M_N )^2 \right]$, from points $P$ (eventually poles)
   located in different Riemann sheets, denoted by the vector ${\bf n}$ as
   introduced in Eq.~(\protect\ref{eq:defLn}), and placed both in the
   first and fourth quadrants. The unitarity cuts are also depicted in
   the figure. The ``distance'' between $S$ and $P$ is obtained by
   the length of the shortest path joining them. This can be achieved 
   after continuous deformation of the paths depicted in the figure,
   i.e. any deformations without intersecting the branch points.}
\label{fig:contour}  
\end{figure}

For four channels, there are a
total number of $2^4 = 16$ Riemann sheets related to unitarity. Now,
$\bar J_0(s)$ and hence $L(s)$ become diagonal matrices in the coupled
channel space, namely 
\be 
{\bf L} (z,{\bf n}) = \left ( \matrix { L_{m_\pi M_N}(z,n_{\pi
N}) & 0 & 0 & 0 \cr 0 & L_{m_\eta M_N}(z,n_{\eta N}) & 0 & 0 \cr 
0 & 0& L_{m_K M_\Lambda}(z,n_{K \Lambda}) & 0 \cr 
0 & 0 & 0 & L_{m_K M_\Sigma}(z,n_{K \Sigma}) } \right ) 
\label{eq:defLn} 
\ee
with $z \in \comp$, ${\bf n} = (n_{\pi N}, n_{\eta N}, n_{K \Lambda},
n_{K \Sigma})$ and the dependence in the masses of the function $L$ is
explicitly given. The first Riemann sheet ($t_I(s)$) corresponds to
the choice ${\bf n} = (0,0,0,0)$. As mentioned above, poles can only
occur in any of the remaining 15 Riemann sheets. The closer the
position of the pole to the scattering region (upper lip of the first
Riemann sheet) the bigger is the influence on the scattering
amplitude. All Riemann sheets can be reached continuously from the
first one by looping around the appropriate branch points. From this
point of view, close means proximity following a continuous path. Thus,
for the region $ (m_\pi+M_N)^2 < s < (m_\eta + M_N)^2$ the poles of
the $(1,0,0,0)-$Riemann sheet located in the fourth quadrant and with
values of ${\rm Re}z$ belonging to the above interval, are expected
to have the biggest influence on the scattering amplitude, as it is
illustrated in Fig.~\ref{fig:contour}.  Similarly,
for the region $ (m_\eta+M_N)^2 < s < (m_K + M_\Lambda)^2$ the poles
of $ (1,1,0,0)-$Riemann sheet located in the fourth quadrant and with
values of ${\rm Re}z$ belonging to the above interval, are expected
to play a crucial role, and so on...
Thus, we define the `` Second Riemann Sheet'' in the relevant fourth
quadrant ($ t_{II}(s)$) as that which is obtained by continuity across
each of the four unitarity cuts. It is obtained using for the diagonal
matrix ${\bar J_0(s)}$ the following function\footnote{Though, each of
the functions ${\bf L}(z;{\bf n})$ are analytical in the complex
plane, except for the pertinent unitarity cuts, note that ${\cal
L}_{II}$ so defined, is continuous for real values of $s$, but presents
additional discontinuities out of the real axis.}

\begin{equation}
{\cal L}_{II} (z) =  \left\{\matrix{ {\bf L}(z;1,0,0,0)
 & {\rm if} &   (m_\pi + M_N )^2 <  {\rm Re} (z) < (m_\eta + M_N )^2 \cr
{\bf L}(z;1,1,0,0) & {\rm if} &  (m_\eta + M_N)^2 < {\rm Re} (z) < ( m_K +
 M_\Lambda)^2  \cr
 {\bf L}(z;1,1,1,0) 
 & {\rm if} &  (m_K + M_\Lambda )^2 < {\rm Re} (z) < (m_K + M_\Sigma )^2
 \cr
{\bf L}(z;1,1,1,1) & {\rm if} & (m_K + M_\Sigma )^2  < {\rm Re}(z) 
\phantom{< (m_K + M_\Sigma )^2}}  \right. \label{eq:secondR}
\end{equation}
\begin{figure}
\centerline{
\epsfig{figure=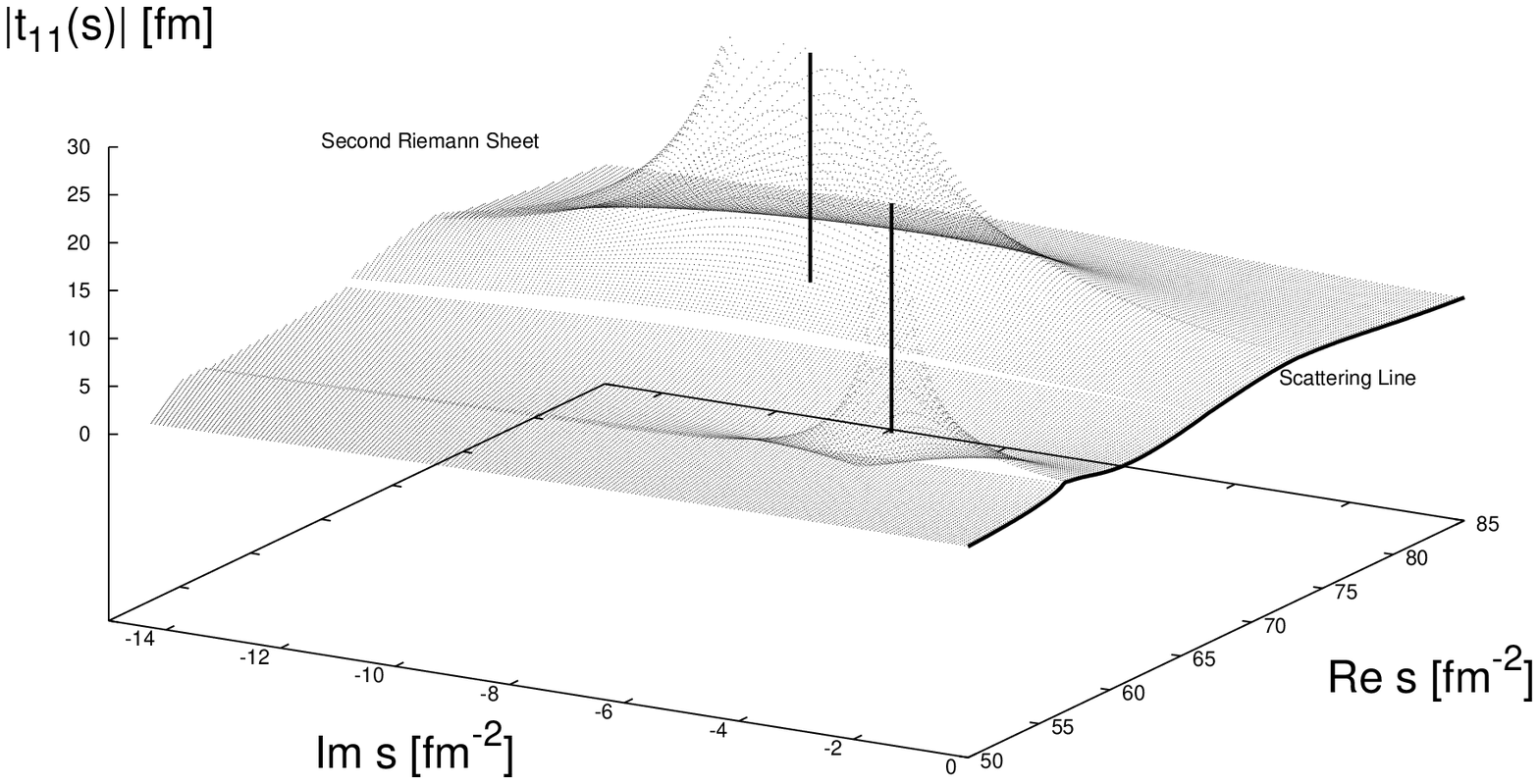,height=9cm,width=17cm}}
\centerline{
\epsfig{figure=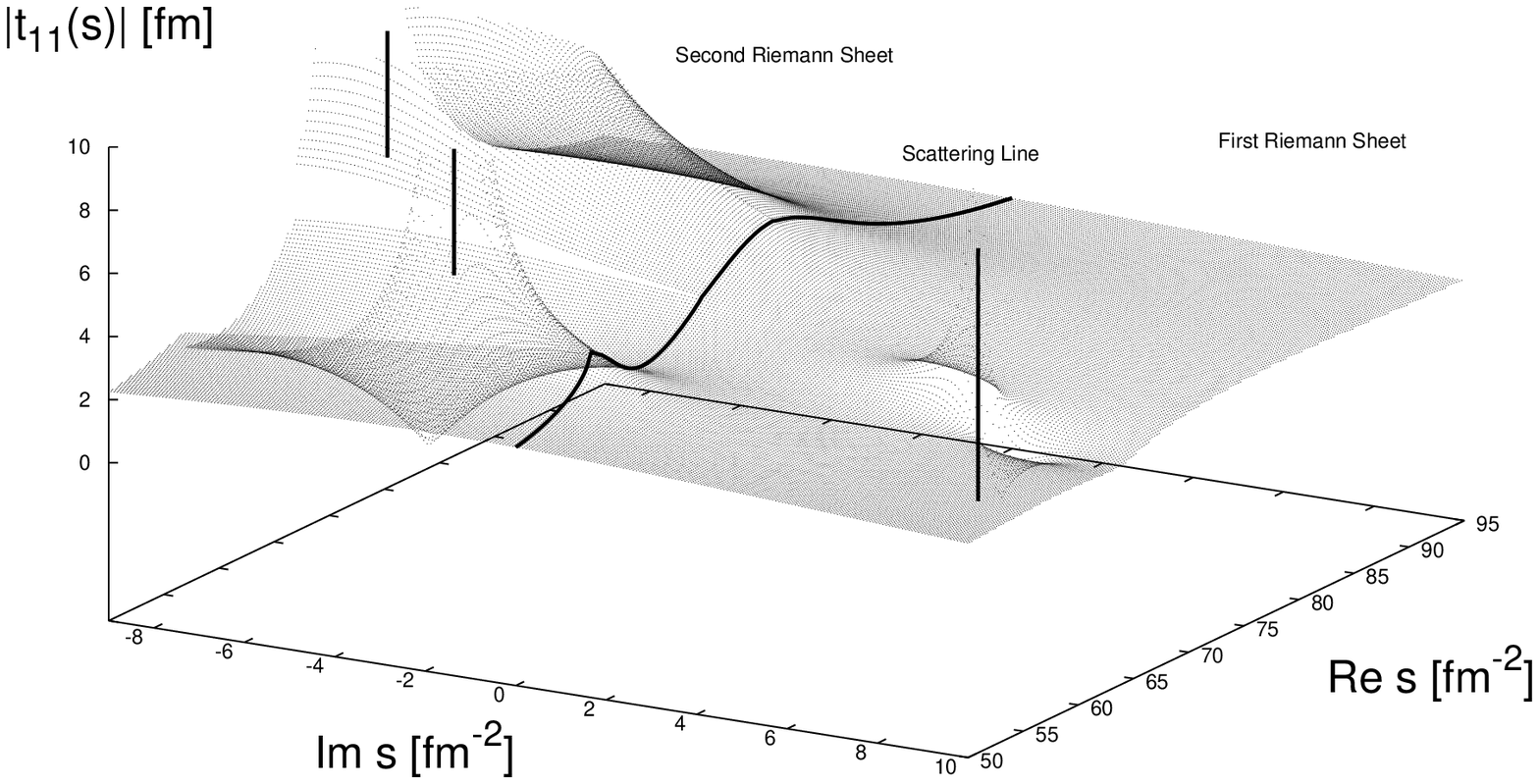,height=9cm,width=17cm}}
\caption{ \footnotesize Modulus of the $\pi N \to \pi N$ element of
the scattering amplitude $t(s)$,
defined in Eq.~(\protect\ref{eq:defts}), in the $s-$complex plane. In
both plots, vertical lines indicate the position of the poles. Top
panel: Fourth quadrant of the ``Second Riemann Sheet'', as defined in
Eq.~(\protect\ref{eq:secondR}), and the physical scattering line. The
two observed poles are identified to be the $S_{11}-$ $N$(1535) and
$-N$(1650) resonances as it is  discussed in the main text. Bottom
panel: Fourth quadrant of the ``Second Riemann Sheet'' and the first
quadrant of the first (physical) Riemann sheet. Besides the two poles
already appearing in the top panel, there is a third one. Though it
is  unphysical because it appears in the physical sheet out of the real
axis, it does not influence the scattering line as the plot clearly 
shows. }
\label{fig:riemman}
\end{figure}
In Fig.~\ref{fig:riemman} we show the absolute value of the $t_{11}(s)$
element of the scattering amplitude both for the fourth quadrant of
the ``Second Riemann Sheet'' and the first quadrant of the first
(physical) Riemann sheet. The physical scattering takes place in the
scattering line in the plots (upper lip of unitarity cut of the first
Riemann sheet). The positions of the two poles in the ``Second Riemann
Sheet'' are ($s = M_R^2 - {\rm i} M_R \Gamma_R$):
\begin{eqnarray}
{\rm First~~\phantom{a} Pole:~~} && M_R = 1496.5 \pm 0.4 \qquad  \Gamma_R
= 83.3 \pm 0.7 \\
&&\nonumber\\  
{\rm Second~~ Pole:~~} && M_R = 1684.3 \pm 0.7 \qquad  \Gamma_R = 194.3 \pm
0.8  
\end{eqnarray}
where all units are given in MeV and errors have been transported from
those in the best fit parameters (Eq.~(\ref{eq:lecs})), taking into
account the existing statistical correlations, through a Monte--Carlo
simulation. These poles are resonances and can be identified to be the
$S_{11}-$ $N$(1535) and $-N$(1650) ones which, according to
Ref.~\cite{PDG98} (PDG), are located at
\begin{eqnarray}
 N(1535):~~~ && M_R = 1505\pm 10 \qquad  \Gamma_R
= 170 \pm 80 \\
&&\nonumber\\  
 N(1650):~~~ && M_R = 1660 \pm 20 \qquad  \Gamma_R = 160 \pm 10  
\end{eqnarray}
where again units are in MeV and we quote data from position of the
poles which are slightly different to those deduced from a
Breit-Wigner fit. The agreement of our predictions and the PDG ones is
satisfactory. Our calculated width for the $N(1535)$ / $N(1650)$ turns
out to be smaller/larger than the experimental one, in great part,
because its mass is slightly smaller/larger than the data and hence
the available phase space for the decay decreases/increases. Besides,
the inclusion of the three body channel ($\pi \pi N$) would influence
both the widths and the masses of the resonances.

Residues at the poles depend on the examined channel, because they
determine the coupling of each of the channels to the
resonances. Thus, from the results shown in Fig.~(\ref{fig:riemman})
one could predict the coupling of the $N(1535)$ and $N(1650)$
resonances to $\pi N$. A detailed study of the couplings of these
resonances to all channels, not only $\pi N$, is presently under way.

On the other hand, there is a unphysical pole in the physical (first)
Riemann sheet. It is located at ($s = M^2 + {\rm i} M \Gamma$) $M
\approx 1582 $ MeV and $\Gamma \approx 166$ MeV and it violates the
Mandelstam's hypothesis of maximal analyticity~\cite{Ma58}. This
unphysical pole appears because we have truncated the iterated
potential to solve the BSE. However, as can be seen in the plots, the
two poles in ``Second Riemann Sheet'' have a much larger influence on
the physical scattering than the spurious (unphysical) one. Thus, the
influence of this unphysical pole may be disregarded\footnote{Due to
Schwartz's Reflection Principle there is also a pole in the fourth
quadrant of the first Riemann sheet (complex conjugated of that given
above) which influence is even more negligible than that of the first
quadrant. This is because it is placed at a substantially larger distance of
the upper lip of the unitarity cut. Existence of other complex
conjugated poles, both in the first Riemann sheet or in any of the
unphysical sheets, is not precluded, but from Fig.~(\ref{fig:riemman})
we infer that their influence in the scattering is not significant.}.

\section{Conclusions}
\label{sec:concl}

 In this paper we have developed a Bethe-Salpeter formalism to study
 $s-$wave and $T=1/2$ meson-baryon scattering up to almost 2 GeV. We
 work on a four dimensional two body channel space and the kernel of
 the BSE takes into account chiral symmetry constraints as deduced
 from the corresponding effective Lagrangian. At lowest order in the
 chiral expansion for the potential, an analytical explicit solution
 is found which manifestly complies with multiple channel
 unitarity. Among the several issues which can be explored using the
 present formalism, we have focused our attention on $\pi N$ elastic
 scattering (phase shifts and inelasticities), and the measured
 inelastic cross sections, being the agreement with experiment
 rather good. Besides, some predictions for other cross sections, not
 yet measured, have been also given.  We have undertaken a careful
 discussion on the analytical structure and continuation of the
 scattering matrix amplitude to the complex $s-$plane, which becomes
 mandatory in order to extract the location of the $S_{11}$ resonances. We
 have searched for poles in the ``Second Riemann Sheet'' and compared
 both masses and widths to data. The agreement is also quite
 satisfactory.  Thus, and despite of having neglected the three body
 production channel $\pi\pi N$, we provide a rather successful
 description of the $s-$wave and $T=1/2$ meson ($\pi, \eta, K$)--baryon
 (N,$\Lambda$, $\Sigma$) scattering up to almost 2 GeV in the
 strangeness zero channel. Couplings of the $N(1535)$ and $N(1650)$
 resonances to the different open meson--baryon channels can be
 obtained from our amplitudes and it will be discussed
 elsewhere. 
 
 Nevertheless our calculation has some obvious limitations and hence
 it might be improved. Besides the inclusion of the $\pi\pi N$
 channel\footnote{In our point of view, it is a highly non trivial
 task to find a solution of the BSE including a three body
 intermediate state exactly complying to three body unitarity. In some
 cases, for instance for elastic $\pi N \to \pi N$, meson $\eta$
 production, ...., some insight might be obtained by treating
 perturbatively the process, as our results suggest.}, one should
 consider the inclusion of higher order terms in the two particle
 irreducible matrix amplitude (potential) which would lead to a
 realistic predictions for higher partial waves in the $T=1/2$
 channel. Understanding the free parameters of our model and their
 numerical values presented in Eq.~(\ref{eq:lecs}), in terms of the
 LEC's appearing in the higher chiral Lagrangian order pieces, either
 within the HBChPT formalism~\cite{fm00} or in the fully covariant
 framework recently developed in Ref.~\cite{bl01}, would be obviously
 desirable. As we have shown in Appendix~\ref{sec:matching}, such a
 task would, at least, require to know the perturbative $1/f^6$ order
 to take into account the likely important $\eta N-$channel effects,
 which are effectively incorporated in the BSE scheme presented in
 this work.

 Given the phenomenological success of the presented framework, it
 seems natural to extend it to other non zero strangeness channels,
 for a recent overview of related aspects see for instance
 Ref.~\cite{ramos01}, or to the study of meson photo-production
 processes.

\section*{Acknowledgments}
We warmly thank C. Garcia-Recio, E. Oset, A. Parre\~no, J.R. Pel\'aez
and A. Ramos for useful discussions. This research was supported by
DGES under contract PB98-1367 and by the Junta de Andalucia.

\appendix

\section{ Basic Integrals} \label{sec:app2}

We display the explicit expressions for the loop integrals used in
this paper. The basic integrals appearing in the solution of the BSE
are:
\begin{eqnarray}
 J( \P ) &=& {\rm i} \int { d^4 q \over (2\pi)^4 } {1\over q^2 - \hat m^2}
\,\, {1\over \slashchar{P}-\slashchar{q} -\hat M }  
\\ 
 J_1^R ( \P ) &=&
{\rm i} \int {
d^4 q \over (2\pi)^4 } {1\over q^2 - \hat m^2}\,\, {1\over
\slashchar{P}-\slashchar{q} -\hat M } \slashchar{q} = J( \P ) (\P -
\hat M ) -
\Delta_{\hat m}  
\\ 
 J_1^L ( \P ) &=&{\rm i} \int { d^4 q \over (2\pi)^4 } {1\over q^2 - 
\hat m^2}
\slashchar{q} {1\over \slashchar{P}-\slashchar{q} - \hat M } \,\, = (\P -
\hat M ) J(\P) - \Delta_{\hat m}  
\\ 
 J_2 ( \P ) &=&
{\rm i} \int { d^4 q \over (2\pi)^4 } {1\over
q^2 - \hat m^2} \slashchar{q} {1\over \slashchar{P}-\slashchar{q}
-\hat M }
\slashchar{q} = (\P - \hat M ) J(\P) (\P - \hat M) - (\P- \hat M) 
\Delta_{\hat m} 
\end{eqnarray} 
and the results are obtained from relativistic and translational
invariance requirement in momentum space. Here, $\Delta_{\hat m}$ is a
quadratically divergent integral 
\begin{eqnarray}
\Delta_{\hat m}&=& {\rm i} \int { d^4 q \over (2\pi)^4 } {1\over q^2 -
\hat m^2}
\label{eq:defdelta}
\end{eqnarray}
which would    require renormalization. Besides, the linearly
divergent integral $J(\P) $ can be evaluated yielding:  
\begin{eqnarray}
J(\P ) &=& 
\slashchar{P} \left[ \left( { s-\hat m^2+\hat M^2 \over 2 s} \right) J_0 (s)
+ {\Delta_{\hat m \hat M} \over 2 s} \right] + \hat M J_0 (s) \label{eq:JP}
\\ && \nonumber \\
 \Delta_{\hat m \hat M} &=& \Delta_{\hat m}-\Delta_{\hat M}= 
{\rm i} \int { d^4 q \over (2\pi)^4 } {1\over q^2 -
\hat m^2}-{\rm i} \int { d^4 q \over (2\pi)^4 } {1\over q^2 -
\hat M^2} \label{eq:deltafit}
\end{eqnarray}  
where $\Delta_{\hat m \hat M}$ is quadratically divergent as well and
the logarithmically divergent integral $J_0 (s)$ needs one subtraction
to make it finite. Choosing for definiteness the threshold value
$s=(\hat m+ \hat M)^2$ we get 
\begin{eqnarray}
J_0 (s) &=& {\rm i} \int { d^4 q \over (2\pi)^4 } {1\over q^2 - \hat
m^2}{1\over (P-q)^2 - \hat M^2 }= \bar J_0 (s) + J_0 (s= (\hat m+ \hat
M)^2)
\label{eq:J0mM}
\end{eqnarray}
with $J_0 (s= (\hat m+ \hat M)^2)$ a divergent integral and the finite
function $\bar J_0 (s) $ is given by\footnote{$J_0(s)$ is a diagonal
matrix in the coupled channel space and for simplicity  we
work from now on in a given channel.} 
\begin{eqnarray}
\bar J_0 (s) &=& {1\over (4\pi)^2} \left\{ \left[ {M^2 -m^2 \over s}
-{M -m \over M + m } \right] {\rm ln} {M \over m } + L(s) \right\}
\label{eq:defls} 
\end{eqnarray} 
and for real $s$ and above threshold, $(m+M)^2$, we have
\begin{eqnarray} 
L(s) \equiv L(s+{\rm i} \epsilon ) = {\lambda^{1/2}(s,m^2,M^2) \over s}
\left\{ \log \left[{ 1+ \sqrt{s-s_+ \over s-s_- } \over 1-\sqrt{s-s_+
\over s-s_- } } \right] -i\pi \right\}
\label{eq:L(s)}
\end{eqnarray}  
where have defined the pseudothreshold  and threshold variables as
\begin{eqnarray}
s_- = (m-M)^2 \qquad s_+ = (m+M)^2 
\end{eqnarray} 
respectively, and the logarithm is taken to be real. Note that $L(s_+)
=0$.  For $ s > s_+ $ the imaginary part along the unitarity cut may
be computed directly from the above Eq.~(\ref{eq:L(s)}) or through
Cutkosky's rules,
\begin{eqnarray}
 2 {\rm i} \, {\rm Im} J_0(s) &=& {\rm Disc} J_0 (s) = \left[ J_0
( s + {\rm i} \epsilon )-J_0 ( s - {\rm i} \epsilon ) \right]  \nonumber\\
&=& {\rm i} \int { d^4 q \over (2\pi)^2 } (-2\pi {\rm i} )^2 \delta^+
(q^2-m^2) \delta^+ ((P-q)^2-m^2) \nonumber\\ &=& - 2 {\rm i} \,
 { \lambda^{1/2} (s,m^2, M^2)
\over 16\pi s} \Theta( s-s_+ ) \label{eq:disj0}
\end{eqnarray}  
Up to a $16\pi^2 $ factor the function $L(s)$ has the same
discontinuity as the function $J_0(s)$. Taking into account that we
have to evaluate the function $J_0(s)$ not only for real $s > s_+ $
but also below threshold\footnote{For instance, when calculating the
elastic $\pi N$ scattering, obviously there are values of $s$ below
heavier thresholds $\eta N$, $K\Lambda$, $K \Sigma $.} and in the
second Riemann sheet as well, to look for the position of resonances
in the complex $s-$plane, we give here the analytical continuation of
$L(z)$ used in our calculation. Defining $\rho_{\pm} = | z-s_{\pm} |$
and taking the principal arguments, Arg(...), $\theta_{\pm}$ of
$|z-s_{\pm}|$ to lie in the range $ 0 \le \theta_+ < 2 \pi $ and $-\pi
\le \theta_- < \pi$ respectively we have
\begin{eqnarray}
L(z,n) &=& {(\rho_+ \rho_- )^{1/2} \over z} e^{ {\rm i} ( \theta_+ +
\theta_- + 2 n \pi )/2 } \Big\{ \ln | R(z) | + {\rm i} {\rm Arg} [R(z)] -
2\pi {\rm i} \Big\} \nonumber \\ && \nonumber \\ R(z) &=& { \rho_+^{1/2}
e^{{\rm i} \theta_+ /2} + \rho_-^{1/2} e^{{\rm i} \theta_- /2} e^{{\rm
i} 
n \pi} \over
\rho_+^{1/2} e^{{\rm i} \theta_+ /2} - \rho_-^{1/2} e^{{\rm i}
\theta_- /2} e^{{\rm i} n \pi} }
\label{eq:L(z,n)}
\end{eqnarray}
where ${\rm Arg} [R(z)]$ should be taken in the interval
$[0,2\pi[$. For $n=0$ one gets the first Riemann sheet
$L_I(z)=L(z,n=0)$, which only has a (unitarity) cut along the real
axis $ s_+ \le s < \infty $. When going across the unitarity cut once
we jump into the second Riemann sheet, corresponding to $n=1$,
$L_{II}(z)=L(z,n=1)$. If we loop twice around the threshold branch
point $z=s_+$ we come back to the original Riemann sheet. The second
Riemann sheet has an additional cut along the real axis $-\infty < s <
s_- $ and the following relation holds
\begin{eqnarray}
L_{II} (z) = L_I(z) + 2 \pi {\rm i} { \lambda^\frac12 (z,m^2,M^2)\over
z}
\end{eqnarray} 
where the cuts for $\lambda^\frac12 (z,m^2,M^2)$ go along the real
axis for $-\infty < s < s_- $ and $ s_+ < s < \infty $. The function
is chosen to be real and positive on the upper lip of the second cut,
$ s_+ < s < \infty $ and corresponds $ |\lambda^\frac12 (s,m^2,M^2)|
\equiv \lambda^\frac12 (s+{\rm i} \epsilon ,m^2,M^2)$. The $-2\pi{\rm
i}$ constant appearing in Eq.~(\ref{eq:L(z,n)}) determines the chosen
Riemann sheet of the logarithm and ensures that $L_I(z)$ is purely
real along the real axis below threshold. Note that since $R(z)$ only
vanishes at infinity we never have a chance to cross the cut of the
logarithm and never change log-Riemann sheets.

\section{Derivation of the solution of the BSE}\label{sec:app1} 

Here we show how to derive Eq.~(\ref{eq:t-1dirac}) displayed in the
main text. The ansatz of Eq.~(\ref{eq:ansatz}) reduces the BSE
integral equation Eq.~(\ref{eq:bse}) into a set of linear equations
for the matrix coefficients $a$,$b_R$,$b_L$ and $c$
\begin{eqnarray}
a &=& a [  ( \P - \hat M) J - \Delta_{\hat m} ] {D \over f^2} + b_L ( \P - \hat M ) [  (
     \P -\hat M )J - \Delta_{\hat m} ] { D \over f^2 } \nonumber
\\ b_R &=& b_R [  (\P -\hat M) J
     -\Delta_{\hat m} ] {D \over f^2 } + c ( \P -\hat M) [  (\P-\hat M)
     J-\Delta_{\hat m}] 
{D\over
     f^2} + {D\over f^2} \nonumber\\ b_L &=& a J {D\over f^2} + b_L [ 
     (\P-\hat M) J -\Delta_{\hat m} ] {D\over f^2} + {D\over f^2} \nonumber\\ 
c &=& b_R J
     {D\over f^2} + c [ (\P -\hat M) J - \Delta_{\hat m} ] {D\over f^2}
\end{eqnarray} 
In the above equation $J$ stands for $J(\P)$ defined and evaluated 
in Appendix \ref{sec:app2}. 
The solution of this matrix system is tricky although straightforward.
The main complication arises from the non-commuting character of the
fermion mass matrix $\hat M$ with the coupled channel matrices
$a$,$b_L$,$b_R$, $c$ and $D$. Defining
\begin{eqnarray} 
X &=& a \qquad Y_R = (\P - \hat M) b_R \qquad Y_L = b_L (\P - \hat M)
\nonumber\\ 
\qquad Z &=& (\P - \hat M) c ( \P - \hat M) \qquad G= {1\over f^2}
\left[  (\P- \hat M) J - \Delta_{\hat m} \right] 
\end{eqnarray} 
The set of matrix equations can be written as 
\begin{eqnarray}
X &=& X G D + Y_L G D 
\label{eq:1} 
\\ Y_R &=& Y_R G D + Z G D + { ( \P - \hat M) D \over f^2 } 
\label{eq:2}\\ Y_L &=&
X ( G + \frac{\Delta_{\hat m}}{f^2} )
 {1\over \P - \hat M } D (\P - \hat M) + Y_L G {1\over
\P - \hat M } D (\P - \hat M) + { D (\P - \hat M) \over f^2 }
\label{eq:3} \\ Z &=& Y_R (G+\frac{\Delta_{\hat m}}{f^2}) 
{1\over \P - \hat M} D (\P - \hat
M) + Z G {1\over \P - \hat M} D (\P - \hat M) \label{eq:4}
\end{eqnarray} 
Summing Eqs.~(\ref{eq:1}) and (\ref{eq:2}), and Eqs.~(\ref{eq:3}) and
(\ref{eq:4}), we get after some matrix reshuffling 
\begin{eqnarray}
{(\P - \hat M) \over f^2 } &=& ( X + Y_R ) ( D^{-1} - G ) - (Y_L + Z )
G \label{eq:5} \\ -{(\P - \hat M) \over f^2 } &=& (X+Y_R )(G+
\frac{\Delta_{\hat m}}{f^2})
- (Y_L + Z) \left[ (\P - \hat M)^{-1} D^{-1} (\P - \hat M) - G \right]
\label{eq:6}
\end{eqnarray} 
Subtracting and summing   Eqs.~(\ref{eq:5}) and (\ref{eq:6}) we 
\begin{eqnarray}
(X+Y_R) &=& (Y_L + Z) \left[ (\P - \hat M)^{-1} D^{-1} (\P - \hat M) (
D^{-1} + \frac{\Delta_{\hat m}}{f^2})^{-1} 
\right] \label{eq:7} \\ {2 (\P - \hat M) \over
f^2 } &=& ( X+Y_R ) ( D^{-1} - 2 G - \frac{\Delta_{\hat m}}{f^2} ) + (Y_L + Z )
\left[ (\P - \hat M)^{-1} D^{-1} (\P - \hat M) - 2 G \right]
\label{eq:8} 
\end{eqnarray}  
respectively.  We can then solve for $Z+Y_L$ from Eqs.~(\ref{eq:7})
and (\ref{eq:8}) yielding 
\begin{eqnarray}
(Y_L + Z)^{-1} &=& \Big\{ \left[ (\P - \hat M)^{-1} D^{-1} (\P - \hat M) (
D^{-1} + \frac{\Delta_{\hat m}}{f^2} )^{-1} +1 \right] (-2G)  \cr  &+&  
(\P - \hat M)^{-1} D^{-1} (\P -\hat M) \left[ ( D^{-1} + 
\frac{\Delta_{\hat m}}{f^2})^{-1} 
( D^{-1} - \frac{\Delta_{\hat m}}{f^2}) +1 \right] \Big\} {f^2 \over 2 (\P - \hat M) }  
\label{eq:9} 
\end{eqnarray} 

Using the proportionality relation between $X+Y_R $ and $Y_L + Z$
given by Eq.~(\ref{eq:7}) we obtain the following expression for the
on-shell $t(\P) $ matrix,
\begin{eqnarray}
t(\P) = X + Y_R + Y_L + Z = (Y_L +Z) \left[ (\P - \hat M)^{-1} D^{-1}
(\P - \hat M) ( D^{-1} + \frac{\Delta_{\hat m}}{f^2} )^{-1} +1 \right]
\end{eqnarray} 
Inverting this equation and using Eq.~(\ref{eq:9}) we get finally the
expression given in Eq.~(\ref{eq:t-1dirac}).

\section{Best fit results}\label{sec:app-stat}

The best fit ($\chi^2 / dof = 0.75$) parameters are 

\begin{eqnarray} 
J_{\pi N}&=&0.1897 \pm 0.0004 \nonumber\\
J_{\eta N} &=& 0.6206 \pm 0.0002 \nonumber\\
J_{K \Lambda} &=& -1.227 \pm 0.003 \nonumber\\
J_{K \Sigma} &=& -0.0143 \pm 0.005 \nonumber\\
\Delta_{N,1} /(m_\pi+M_N)^2 &=& 0.776 \pm 0.002 \nonumber\\
\Delta_{N,2}/(m_\eta+M_N)^2 &=& 1.8375 \pm 0.0004 \nonumber\\
\Delta_{\Lambda}/(m_K+M_\Lambda)^2  &=& -2.923 \pm 0.008 \label{eq:lecs}\\
\Delta_{\Sigma} /(m_K+M_\Sigma)^2 &=& 1.000 \pm 0.017 \nonumber\\ 
\Delta_{\pi} /(m_\pi+M_N)^2&=& -0.0123 \pm 0.0003 \nonumber\\
\Delta_{\eta } /(m_\eta+M_N)^2&=& -0.1560 \pm 0.0002 \nonumber\\
\Delta_{K,1 } /(m_K+M_\Lambda)^2 &=& -0.006324 \pm 0.000003 \nonumber\\ 
\Delta_{K,2}  /(m_K+M_\Sigma)^2 &=& 0.001128 \pm 0.000003 \nonumber
\end{eqnarray} 
The correlation matrix, defined as usual,
\begin{eqnarray}
r_{ij} &=& \langle  x_i x_j \rangle \nonumber\\
&&\nonumber\\
x_i &=& \frac{c_i -\langle c_i \rangle }{\sqrt{ \langle c_i^2\rangle -
\langle c_i \rangle^2}}\nonumber\\
&&\nonumber\\
\langle f(c_1,\dots,c_n) \rangle & = & 
\frac{1}{N} \sum_{\alpha=1}^N f(c_{1,\alpha},\dots,c_{n,\alpha})
\end{eqnarray}
being $c_i$ any of the 12 parameters: $J's$ and $\Delta 's$, turns out
to be:

\begin{eqnarray}
{\scriptsize
\left(
\matrix{
  \phantom{-}1.00&     &	  &     &     &     &     &     &     &     &     &\cr
 -0.62& \phantom{-}1.00&     &     &     &     &     &     &     &     &     &\cr
  \phantom{-}0.35& \phantom{-}0.21& \phantom{-}1.00&     &     &     &     &     &     &     &     &\cr
  \phantom{-}0.28&-0.28& \phantom{-}0.48& \phantom{-}1.00&     &     &     &     &     &     &     &\cr
  \phantom{-}0.67&-0.22& \phantom{-}0.57& \phantom{-}0.09& \phantom{-}1.00&     &     &     &     &     &     &\cr
  \phantom{-}0.63&-0.35&-0.18&-0.15& \phantom{-}0.00& \phantom{-}1.00&     &     &     &     &     &\cr
  \phantom{-}0.32& \phantom{-}0.24& \phantom{-}0.97& \phantom{-}0.42& \phantom{-}0.49&-0.10& \phantom{-}1.00&     &     &     &     &\cr
  \phantom{-}0.36&-0.30& \phantom{-}0.51& \phantom{-}0.99& \phantom{-}0.14&-0.09& \phantom{-}0.43& \phantom{-}1.00&     &     &     &\cr
  \phantom{-}0.50&-0.51&-0.30&-0.08&-0.21& \phantom{-}0.91&-0.25&-0.04& \phantom{-}1.00&     &     &\cr
 -0.51& \phantom{-}0.48&-0.48&-0.62&-0.71& \phantom{-}0.30&-0.33&-0.65& \phantom{-}0.30& \phantom{-}1.00&     &\cr     
 \phantom{-} 0.10&-0.37& \phantom{-}0.12& \phantom{-}0.20& \phantom{-}0.56&-0.59& \phantom{-}0.01& \phantom{-}0.18&-0.55&-0.77& \phantom{-}1.00&\cr
  \phantom{-}0.19&-0.03& \phantom{-}0.54& \phantom{-}0.48& \phantom{-}0.67&-0.56& \phantom{-}0.42& \phantom{-}0.48&-0.69&-0.84& \phantom{-}0.72&
\phantom{-}1.00} \right)}
\end{eqnarray}

The large correlations (0.97 and 0.99) between the pairs $J_{K
\Lambda}-\Delta_{\Lambda}$ and $J_{K \Sigma}-\Delta_{\Sigma}$ can be
understood by looking at Eq.~(\ref{eq:s11-amp}) and taking into
account the smallness of the parameters $\Delta_{K,1 }$  and
$\Delta_{K,2 }$ respectively.

\section{The heavy baryon limit and HBChPT} 

\label{sec:matching}

It is in principle of theoretical and phenomenological interest the
study of the heavy baryon limit of the BSE amplitude given in
Eq.~(\ref{eq:s11-amp}). In the static limit, baryons behave like fixed
sources, and consequently the two particle problem should reduce to a
one particle scattering problem (in our case of meson-baryon
scattering it would correspond to a Klein-Gordon equation with a
spin-dependent potential). It has been known that the BSE has some
difficulties in reproducing this heavy-light limit in certain
situations (ladder approximation to one boson
exchange~\cite{gross82}). We show below that our amplitude has a
correct one particle limit, due to the fact the $s-$wave interaction
is of the contact type. If, in addition to a heavy baryon expansion, a
chiral expansion in powers of $1/f^2$ is carried out, we should
recover in this double expansion some form of the results found in
Ref.~\cite{Mo98,Fettes98,ej00c} within HBChPT for the elastic $\pi N$
scattering amplitude. The heavy baryon expansion may be taken by
making the baryon masses $\hat M\to \infty$ but keeping the meson masses,
$\hat m$, and the meson momentum, $q$, finite. On the other hand, baryon
mass splittings must be considered higher order effects, see
e.g. \cite{Pich95}, so that we take the mass matrix
\begin{eqnarray}
\hat M &=& M_B + \Delta \hat M 
\end{eqnarray} 
with $M_B \to \infty$ the common mass of the baryon octect which is
proportional to the identity matrix. Accordingly, in the $\pi N$
elastic channel we take
\begin{eqnarray}
\sqrt{s} = E+\omega = M_N + \omega + \frac{\omega^2-m_\pi^2}{2M_N} +
\cdots
\end{eqnarray}
where $M_N = M_B + \Delta M_N $.  In this appendix we match our
amplitude to the HBChPT third order results of Ref.~\cite{ej00c} based
on the previous analysis of Refs.~\cite{Mo98,Fettes98}. The heavy
baryon expansion can be directly done for explicit variables, such as
masses and CM energy $\sqrt{s}$. The constants $J_{\pi N}, \dots,
\Delta_{N,1} , \dots, \Delta_\pi, \dots, $ given by our numerical fit
in Eq.~(\ref{eq:lecs}), also might have a baryon mass dependence. Such
a dependence should lead to some changes in the heavy baryon expansion
which cannot be easily determined. In addition, given the
non-perturbative nature of our solution and the fact that many aspects
of the coupled channel meson-baryon data in the $S_{11}$--strangeness
zero channel are described after unitarization up to energies as high
as $\sqrt{s}= 2{\rm GeV}$, it seems obvious that the parameters of
Eq.~(\ref{eq:lecs}) also incorporate higher order effects in the 
chiral expansion.

\subsection{Static limit} 

It is convenient to do the study in terms of the inverse coupled
channel matrix amplitude, $t^{-1} (s)$, given by
Eq.~(\ref{eq:s11-amp}). From the expression of the one-loop integral
Eq.~(\ref{eq:defls}) and Eq.~(\ref{eq:L(s)}) we get to leading order
\begin{eqnarray}
M \bar J_0 (s,m,M) |_{\sqrt{s}=\sqrt{M_N^2+\omega^2-m_\pi^2}+\omega}=
\frac1{16 \pi^2} \log \left( \frac{M^2}{m^2} \right) (m-\omega)+\bar
K_m (\omega) + {\cal O}\left( \frac1M \right)
\end{eqnarray} 
The subtraction point for the HBChPT integrals is $\omega=m$, $\bar
K_m (m) =0$, and their explicit expression is 
\begin{eqnarray}
\bar K_m (\omega) = \frac1{8 \pi^2 } \cases{ -\sqrt{\omega^2-m^2} \, {\rm
arccosh} \left(-\frac{\omega}m \right) \hfill \omega  < -m \cr
+\sqrt{m^2-\omega^2} \, {\rm arccos} \left(-\frac{\omega}m \right)
\hfill \omega^2 < m^2 \cr +\sqrt{\omega^2-m^2} \left[ {\rm arccosh} \left(
\frac{\omega}m \right) - {\rm i} \pi \right] \hfill \omega > m }
\end{eqnarray} 
Thus, one obtains for the unsubtracted integral 
\begin{eqnarray}
M J_0 (s,m,M) |_{\sqrt{s} = \sqrt{M_N^2+\omega^2-m_\pi^2}+\omega} &=&
K_{mM} (\omega) + {\cal O}\left( \frac1M \right) \nonumber \\ &=& M
J_{mM}^0 + \frac1{16 \pi^2} \log \frac{M^2}{m^2} (m-\omega)+\bar K_m
(\omega) + {\cal O}\left( \frac1M \right)
\end{eqnarray} 
where the HBChPT unsubtracted integrals $K_{mM}(\omega) $, fulfilling
$K_{mM} (m)= M J_{mM}$ and the heavy baryon approximation of the
subtraction constant defined through Eq.~(\ref{eq:J0mM}), 
\begin{eqnarray} 
J_0 (s=(m+M)^2 ) &=& J_{mM}^0 \left\{ 1 + {\cal O} \left( \frac1M
\right)\right \}
\end{eqnarray} 
have been introduced. In the static limit we obtain from
Eqs.~(\ref{eq:deff0}) and~(\ref{eq:s11-amp})  
($f(\omega) \to -t(s)/(4\pi)$ ) 
\begin{eqnarray}
 f (\omega )^{-1} &=& 8 \pi \left[ \bar K_{\hat m} (\omega) +
\frac1{16 \pi^2} \ln \frac{\hat M^2}{\hat m^2} (\hat m-\omega)+ \hat M
J_{\hat m,\hat M}^0 +\frac{\Delta_{\hat m \hat M}^0}{4 \hat M} \right]
\nonumber \\ &-& \frac{4\pi}\omega\left\{ \Delta^0_{\hat m} - \left[
\frac2{f^2} D + \frac1{f^4}D \Delta^0_{\hat m} D \right]^{-1} \right\}
\label{eq:static}
\end{eqnarray} 
with $\bar K_{\hat m} (\omega) = {\rm Diag} ( \bar K_{\pi}
(\omega), \bar K_{\eta} (\omega) , \bar K_{K} (\omega), \bar K_{K}
(\omega) ) $ and the heavy baryon approximation of the subtraction 
constants are defined by means of the expansion
\begin{eqnarray} 
\Delta_{\hat m} &=& \Delta_{\hat m}^0 \left\{ 1 + {\cal O} \left(
\frac1M \right)\right\} \nonumber \\ \Delta_{\hat m \hat M} &=& 
\Delta_{\hat m \hat M}^0 \left\{ 1 + {\cal O} \left( \frac1M \right)\right\} 
\nonumber
\end{eqnarray} 
The Eq.~(\ref{eq:static}) corresponds, as it should, to a
one particle scattering problem, fulfilling the coupled channel
unitarity condition
\begin{eqnarray} 
{\rm Im}  f (\omega)^{-1} = -\sqrt{\omega^2-\hat m^2} \theta (
\omega - \hat m) 
\end{eqnarray} 
The pole in Eq.~(\ref{eq:static}) for the inverse amplitude is a
static limit reminiscent from the baryonic Adler zero, $\sqrt{s}-\hat
M=0$, of the lowest order potential.  The constant combination
appearing in the inverse amplitude, Eq.~(\ref{eq:static}), $ \hat M
J_{\hat m\hat M}^0 + \Delta_{\hat m \hat M}^0 /4 \hat M$ should go to
some definite value in the static limit, $M \to \infty$. In case it
would diverge, the scattering amplitude would become trivial. We may
try to estimate these constants using the numerical values obtained in
the $\chi^2-$fit carried out in this work and given in
Eq.~(\ref{eq:lecs}). We get
\begin{eqnarray} 
 M_N J_{\pi N}+ \frac{\Delta_{\pi N}}{4 M_N } &=& -0.47 m_\pi 
\nonumber \\  M_N J_{\eta N}+ \frac{\Delta_{\eta N}}{4 M_N } &=&
-1.08 m_\eta  \nonumber \\  M_\Lambda J_{K \Lambda}+
\frac{\Delta_{K \Lambda}}{4 M_\Lambda} &=& +0.67 m_K  \nonumber \\
 M_\Sigma J_{K \Sigma}+ \frac{\Delta_{K \Sigma}}{4 M_\Sigma} &=&
-1.24 m_K 
\end{eqnarray} 
in units of the relevant pseudoscalar meson masses. Though the
numerical values used for the subtraction constants  contain
higher order effects in the heavy baryon expansion, we see that there
is indeed some trend to cancellation, because the $J's$ and the
$\Delta 's$ contributions have opposite signs in the first three cases,
and $J_{K \Sigma}$ is very small. Moreover, the constants do not seem
to attain unnaturally large values, although it is hard to say which
should be an accurate appropriate value. 

\subsection{Chiral and heavy baryon expansion} 

Expanding Eq.~(\ref{eq:s11-amp}) in powers of $1/f^2 $ we get
\begin{eqnarray}
t(s) &=& t_2(s) + t_4 (s) + \cdots 
\end{eqnarray} 
where 
\begin{eqnarray}
t_2 (s) &=& \frac1{f^2} \left\{ \sqrt{s}-\hat M, D \right\} \\ t_4 (s)
&=& \frac1{f^4} \left( \sqrt{s}-\hat M \right) D \frac{\Delta_{\hat
m}}{\sqrt{s}-\hat M} D \left( \sqrt{s}-\hat M \right) \\ &+&
\frac1{f^4} \left\{ \sqrt{s}-\hat M, D \right\} \left( {
(\sqrt{s}+\hat M)^2-\hat m^2 \over 2 \sqrt{s}} \hat J_0 (s) -
{\Delta_{\hat m} \over \sqrt{s}-\hat M} + \frac{\Delta_{\hat m \hat
M}}{2 \sqrt{s}} \right) \left\{ \sqrt{s}-\hat M, D \right\}
\nonumber
\end{eqnarray} 
In the heavy baryon limit we get for the elastic $\pi N \to \pi N $
amplitude in the $S_{11}$ channel
\begin{eqnarray}
f (\omega) &=& f_2(\omega) + f_4 (\omega) + \cdots 
\end{eqnarray}
with 
\begin{eqnarray}
f_2 (\omega) = &+&\frac{\omega}{4\pi f_\pi^2}
-\frac{m_\pi^2+\omega^2}{8\pi f_\pi^2 M_N}+
\frac{\omega(3m_\pi^2+\omega^2)}{16\pi f_\pi^2 M_N^2} + {\cal O}\left(
\frac1{M_N^3 f^2} \right) \\ f_4 (\omega) = &-& \frac{\omega^2}{64 \pi
f_\pi^2 } \left[ \frac{16}{f_\pi^2} \left( 2 K_{\pi N} (\omega) +
{\Delta_{\pi N}\over 2 M_N } \right) + \frac9{f_K^2 } \left( 2
K_{K\Lambda} (\omega) + {\Delta_{K \Lambda}\over 2 M_\Lambda } \right) +
\frac1{f_K^2} \left( 2 K_{K \Sigma} (\omega) + {\Delta_{K \Sigma}\over
2 M_\Sigma } \right) \right] \nonumber \\ &+& \frac{3\omega}{ 256 \pi
f_\pi^2} \left[ \frac{16}{f_\pi^2} \Delta_{\pi}+\frac{9}{f_K^2}
\Delta_{K,1} + \frac1{f_K^2}\Delta_{K,2} \right]+ {\cal O}\left(
\frac1{M_N f^4} \right)
\label{eq:f2f4}
\end{eqnarray}
In the  region\footnote{Note, that the $\eta N$ channel appears
at order $1/f^6$.}   $m_\pi \le \omega \le m_K $ only the $K_{\pi N}
(\omega)$ has an imaginary part, to comply with perturbative elastic
unitarity, whereas $K_{K \Lambda} (\omega) $ and $K_{K \Sigma}
(\omega) $ are purely real. To write down this expression we have
considered the prescription $ D/f^2 \to \hat f^{-1} D \hat f^{-1} $
given in Eq.~(\ref{eq:f-presc}). At threshold, $\omega = m_\pi$, the
$\pi N $ scattering length in this channel reads 
\begin{eqnarray}
a_{\pi N} &=&\frac{m_\pi}{4\pi f_\pi^2}-\frac{m_\pi^2}{4\pi f_\pi^2
M_N}+ \frac{m_\pi^3}{4\pi f_\pi^2 M_N^2} -\frac{m_\pi^2}{64 \pi
f_\pi^2 } \left\{ \frac{16}{f_\pi^2} \left[ 2 M_N J_{\pi N}^0 +
{\Delta_{\pi N}^0 \over 2 M_N } \right]\right. \nonumber \\ &+& \frac9{f_K^2
} \left[ 2 M_\Lambda J_{K\Lambda}^0 + {\Delta_{K \Lambda}^0\over 2
M_\Lambda } + \frac1{4\pi^2} \left\{ (m_\pi - m_K)\log
\frac{M_\Lambda}{m_K} + \sqrt{m_K^2-m_\pi^2}
{\rm~arccos}\left(-\frac{m_\pi}{m_K} \right) \right\} \right] \nonumber \\
&+& \left. \frac1{f_K^2} \left[ 2 M_\Sigma J_{K \Sigma}^0 + {\Delta_{K
\Sigma}^0\over 2 M_\Sigma } + \frac1{4\pi^2} \left\{ (m_\pi-m_K)\log
\frac{M_\Sigma}{m_K} + \sqrt{m_K^2-m_\pi^2 }
{\rm ~arccos}\left(-\frac{m_\pi}{m_K} \right) \right\} \right] \right \}
\nonumber \\  &+&\frac{3 m_\pi }{ 256 \pi f_\pi^2} \left[
\frac{16}{f_\pi^2} \Delta_{\pi}^0+\frac{9}{f_K^2} \Delta_{K,1}^0 +
\frac1{f_K^2}\Delta_{K,2}^0 \right] + {\cal O}\left( \frac1{f^2
M_N^3}, \frac1{M_N f^4} , \frac1{f^6} \right)
\label{eq:a-piN}
\end{eqnarray}
From Eq.~(\ref{eq:f2f4}) and Eq.~(\ref{eq:a-piN}) above it is clear
that there is no contribution from the $\eta N$ channel to this
$1/f^4$ order of approximation. This is a direct consequence of the
structure of the coupled channel matrix $D$ given by
Eq.~(\ref{d-matrix}) since the corresponding $\pi N \to \eta N $
transition matrix element vanishes in the Born approximation. For this
reason, the channel $\eta N$ starts contributing at order
$1/f^6$. This situation is unexpected, because on general grounds we
expect the $\eta N$ channel to be more important at low energies than
the $K \Lambda $ and $K \Sigma$ channels since their thresholds lie
at higher energies. Using the numerical values of the coefficients
obtained from the fit of Eq.~(\ref{eq:lecs}) to estimate the scattering
length, Eq.~(\ref{eq:a-piN}), we get
\begin{eqnarray}
a_{\pi N} = \underbrace{0.22}_{1/f^2} +
\underbrace{\overbrace{0.22}^{\pi N}- \overbrace{1.06}^{K
\Lambda}+\overbrace{0.18}^{K \Sigma}}_{1/f^4} =-0.43 \,~ {\rm fm}
\end{eqnarray}
which should be compared to the one of Eq.~(\ref{eq:apin}) obtained
from the full amplitude, $0.179~ {\rm fm}$. Obviously, the different
values should be attributed to non negligible higher order effects,
which in particular include $\eta N$ contributions, and higher order
corrections to the $J^0$ and $\Delta^0 $ coefficients. This is very
reassuring because the full BSE amplitude, Eq.~(\ref{eq:s11-amp}),
besides restoring unitarity automatically includes all orders in the
chiral expansion.

\def\Ln{{\rm arccosh} \left( \frac{\omega}m_\pi \right)} \def\Lnn{\log
\left( \frac{\omega^2}{m_\pi^2} \right) } \def\Atan{{\rm arctan}
\left( \frac{\sqrt{\omega^2-m_\pi^2}}{m_\pi} \right)}

\subsection{Matching to HBChPT}

The $S_{11}$ partial wave amplitude deduced from the work of
Ref.~\cite{ej00c} based in HBChPT to third order \cite{Mo98,Fettes98}
reads, after straightforward angular integration, 
\begin{eqnarray}
f_2 (\omega)&=& \frac{\omega}{4 f_\pi^2\pi} + \frac{m_\pi^4 g_A^2 +
\omega^2 m_\pi^2 \left[- 6  - 48 a_3 + 4 g_A^2 \right] + \omega^4
\left[ - 6 + 48 (a_1+a_2) -5 g_A^2 \right] }{48 f_\pi^2
M_N\pi \omega^2 } \nonumber \\ &+& \frac{ m_\pi^6 g_A^2 -
g_A^2\omega^2 m_\pi^4 + \omega^4 m_\pi^2 \left[6 - 48(a_2-2 a_3) - 7
g_A^2 \right] + \omega^6  \left[ 6 - 48 a_1  + 7 g_A^2 \right]}{48
f^2 M_N^2\pi \omega^3 } \\
f_4 (\omega) &=& -\frac{\omega^2 \bar K_\pi(\omega)}{2 \pi f_\pi^4} 
+ \frac1{5760 f_\pi^4\pi^3}\Big( -102 m_\pi^3 g_A^2\pi + \omega
m_\pi^2 \left[ - 130 + 1440 \tilde b_6 - 230 g_A^2 \right] \nonumber
\\  && + 144 g_A^2\pi m_\pi \omega^2 + \omega^3 \left[ 115 + 720 ( \tilde b_1+
\tilde b_2+ \tilde b_3 ) - 205 g_A^2 \right] \Big) \nonumber \\ &+& \frac{13
g_A^2 m_\pi^5 }{3840 f_\pi^4\pi^2 (m_\pi^2- \omega^2)} \Lnn \nonumber
\\ &+& \frac{(1 + g_A^2) m_\pi^4 \omega}{128 f_\pi^4\pi^3
(\omega^2-m_\pi^2)} \left[\Ln\right]^2 \nonumber \\ &+& \frac{
\omega^2 \left[ 3(1+ g_A^2) m_\pi^2 + 2 \omega^2 ( 1 + 5g_A^2 )
\right] } { 192 f_\pi^4\pi^3 \sqrt{\omega^2-m_\pi^2}} \Ln \nonumber \\
&+& \frac{ g_A^2 (13 m_\pi^4 - 46 \omega^2 m_\pi^2 + 48
\omega^4)}{1920 f_\pi^4\pi^2 \sqrt{\omega^2-m_\pi^2}} \Atan 
\label{eq:fmoj}
\end{eqnarray} 
There is no unique way to match the low energy chiral expansion of the
coupled channel BSE amplitude, Eq.~(\ref{eq:f2f4}), to the third order
HBChPT calculation of Refs.~\cite{Mo98,Fettes98},
Eq.~(\ref{eq:fmoj}). The analytical structure is different besides the
elastic unitarity cut at $\omega=m_\pi$ which turns out to
coincide. Indeed, while the former presents the inelastic unitarity
cuts for the considered $K \Lambda $ and $K \Sigma $ coupled channels,
the latter includes perturbatively the left hand cut at
$\omega=0$. Obviously, any particular choice of the matching point
generates a specific set of low energy constants. After explicitly
separating the elastic unitarity correction of both amplitudes, it
seems reasonable to do the matching of the remaining pieces in a
polynomial expansion around the elastic threshold point,
$\omega=m_\pi$, since neither inelastic unitarity cuts nor the left
cut are expected to be crucial at that point. Instead, we expect both
amplitudes to provide a sensible approximation. Also, direct
inspection of Eq.~(\ref{eq:f2f4}) and Eq.~(\ref{eq:fmoj}) reveals that
only some additive combinations among renormalization constants can be
established. In particular, in Eq.~(\ref{eq:f2f4}) there are two
independent combinations of low energy constants. Thus, it proves
sufficient to Taylor expand around $\omega = m_\pi $ up to first
order. Using the numerical values for the input parameters,
Eq.~(\ref{eq:numval}), in Eq.~(\ref{eq:f2f4}) and the numerical values
for the parameters in Ref.~\cite{ej00c} in Eq.~(\ref{eq:fmoj}) the
following identifications hold, (in units of fm)
\begin{eqnarray} 
0.185 &=& 0.175 + 22.8 {\bar \Delta}^0_\pi + 2.51 {\bar
\Delta}^0_{N,1} + 17.3 {\bar \Delta}^0_{K,1} + 1.57 {\bar
\Delta}^0_\Lambda + 2.12 {\bar \Delta}^0_{K,2} + 0.179 {\bar
\Delta}^0_\Sigma \nonumber \\ &-& 7.60 J_{\pi N}^0 - 3.01 J_{K
\Lambda}^0 - 0.357 J_{K \Sigma}^0 \nonumber \\ -0.051 &=& 0.163 + 20.3
{\bar \Delta}^0_\pi + 5.01 {\bar \Delta}^0_{N,1} + 15.7 {\bar
\Delta}^0_{K,1} + 3.14 {\bar \Delta}^0_\Lambda + 1.94 \bar
\Delta^0_{K,2} + 0.359 {\bar \Delta}^0_\Sigma \nonumber \\ &-& 15.2
J_{\pi N}^0 - 6.01 J_{K \Lambda}^0 - 0.714 J_{K \Sigma}^0 \label{eq:matching}
\end{eqnarray} 
where $\bar \Delta_{\hat m}^0 = \Delta_{\hat m}^0 /(m+M)^2 $ and $\bar
\Delta_{\hat M}^0 = \Delta_{\hat M}^0 /(m+M)^2 $ are dimensionless. As
we see, there is a large degree of redundancy when the matching is
performed considering only these low orders of
the expansion.  By using the values of Eq.~(\ref{eq:lecs}), to estimate
the heavy-baryon mass independent parameters appearing in the right
hand side of the two relations established in Eq.~(\ref{eq:matching}),
we obtain $-$0.43 and $-$0.618 respectively. The disagreement, with
respect to the left hand side values, is not completely surprising
because the numerical values used for the subtraction constants
contain higher order effects in the heavy baryon expansion. Besides,
the nominally small differences (${\cal O}(1/M)$) between $\bar
\Delta^0$'s and $J^0$'s and $\bar \Delta$'s and $J$'s might lead to 
significant numerical changes because the factors multiplying these 
constants are large in units of the left hand side values of
Eq.~(\ref{eq:matching}).

\end{document}